# Adaptive Simulated Annealing: A Near-optimal Connection between Sampling and Counting

Daniel Štefankovič[*]    Santosh Vempala[†]    Eric Vigoda[‡]

December 9, 2006

## Abstract

We present a near-optimal reduction from approximately counting the cardinality of a discrete set to approximately sampling elements of the set. An important application of our work is to approximating the partition function $Z$ of a discrete system, such as the Ising model, matchings or colorings of a graph. The typical approach to estimating the partition function $Z(\beta^*)$ at some desired inverse temperature $\beta^*$ is to define a sequence, which we call a *cooling schedule*, $\beta_0 = 0 < \beta_1 < \cdots < \beta_\ell = \beta^*$ where $Z(0)$ is trivial to compute and the ratios $Z(\beta_{i+1})/Z(\beta_i)$ are easy to estimate by sampling from the distribution corresponding to $Z(\beta_i)$. Previous approaches required a cooling schedule of length $O^*(\ln A)$ where $A = Z(0)$, thereby ensuring that each ratio $Z(\beta_{i+1})/Z(\beta_i)$ is bounded. We present a cooling schedule of length $\ell = O^*(\sqrt{\ln A})$.

For well-studied problems such as estimating the partition function of the Ising model, or approximating the number of colorings or matchings of a graph, our cooling schedule is of length $O^*(\sqrt{n})$, which implies an overall savings of $O^*(n)$ in the running time of the approximate counting algorithm (since roughly $\ell$ samples are needed to estimate each ratio).

A similar improvement in the length of the cooling schedule was recently obtained by Lovász and Vempala in the context of estimating the volume of convex bodies. While our reduction is inspired by theirs, the discrete analogue of their result turns out to be significantly more difficult. Whereas a fixed schedule suffices in their setting, we prove that in the discrete setting we need an adaptive schedule, i.e., the schedule depends on $Z$. More precisely, we prove any non-adaptive cooling schedule has length at least $O^*(\ln A)$, and we present an algorithm to find an adaptive schedule of length $O^*(\sqrt{\ln A})$.

## 1 Introduction

This paper explores the intimate connection between counting and sampling problems. By counting problems, we refer to estimating the cardinality of a large set (or its weighted analogue), or in a continuous setting, an integral over a high-dimensional domain. The sampling problem refers to generating samples from a probability distribution over a large set. The well-known connection between counting and sampling is the starting point for popular Markov chain Monte Carlo methods for many counting problems. Some notable examples from computer science are the problems of estimating the volume of a convex body [3, 11] and approximating the permanent of a non-negative matrix [9].

---

[*]Department of Computer Science, University of Rochester, Rochester, NY 14627, and Comenius University, Bratislava. Email: stefanko@cs.rochester.edu

[†]College of Computing, Georgia Institute of Technology, Atlanta GA 30332. Email: vempala@cc.gatech.edu. Supported by the NSF and a Guggenheim fellowship.

[‡]College of Computing, Georgia Institute of Technology, Atlanta GA 30332. Email: vigoda@cc.gatech.edu. Research supported in part by NSF grant CCF-0455666.

In statistical physics, a key computational task is estimating a partition function, which is an example of a counting problem. Evaluations of the partition function yield estimates of thermodynamic quantities of interest, such as the free energy and the specific heat. The corresponding sampling problem is to generate samples from the so-called Gibbs (or Boltzman) distribution.

We present an improved reduction from approximate counting to approximate sampling. These results improve the running time for many counting problems where efficient sampling schemes exist. We present our work in the general framework of partition functions from statistical physics. This framework captures many well-studied models from statistical physics, such as the Ising and Potts models, and also captures many natural combinatorial problems, such as colorings, independent sets, and matchings., For the purpose of this paper we define a (discrete) partition function as follows.

**Definition 1.1.** Let $n \geq 0$ be an integer. Let $a_0, \ldots, a_n$ be non-negative real numbers such that $a_0 \geq 1$. The function
$$Z(\beta) = \sum_{i=0}^{n} a_i e^{-i\beta}$$
is called a *partition function* of degree $n$. Let $A := Z(0)$.

This captures the standard notion of partition functions from statistical physics in the following manner. The quantity $i$ corresponds to the possible values of the Hamiltonian. Then $a_i$ is the number of configurations whose Hamiltonian equals $i$. For instance, in the (ferromagnetic) Ising model on a graph $G = (V, E)$, a configuration is an assignment of $+1$ and $-1$ spins to the vertices. The Hamiltonian of a configuration is the number of edges whose endpoints have different spins. The quantity $\beta$ is referred to as the inverse temperature. The computational goal is to compute $Z(\beta)$ for some choice of $\beta \geq 0$. Note, when $\beta = 0$ the partition function is trivial since $Z(0) = \sum_{i=0}^{n} a_i = 2^{|V|}$. The condition $a_0 \geq 1$ is clearly satisified, in fact, we have $a_0 = 2$ by considering the all $+1$ and the all $-1$ configurations.

The general notion of partition function also captures standard combinatorial counting problems as illustrated by the following example. Let $\Omega$ be the set of all $k$-labelings of a graph $G = (V, E)$ (i. e., labelings of the vertices of $G$ by numbers $\{1, \ldots, k\}$). Given a labeling $\sigma$, let its Hamiltonian $H(\sigma)$ be the number of edges in $E$ that are monochromatic in $\sigma$. Let $\Omega_i$ denote the set of all $k$-labelings of $G$ with $H(\sigma) = i$. Let $a_i = |\Omega_i|$. We would like to compute $Z(\infty) = a_0$, i.e., the number of valid $k$-colorings of $G$. Once again, the case $\beta = 0$ is trivial since we have $Z(0) = k^{|V|}$. The condition $a_0 \geq 1$ simply requires that there is at least one proper $k$-coloring.

The standard approach to compute $Z(\beta)$ is to express it as a telescoping product of ratios of the partition function evaluated at a sequence of $\beta$'s, where the initial $\beta = 0$ is the trivial case. The ratios are approximated using a sampling algorithm. More precisely, consider a set of configurations $\Omega$ which can be partitioned as $\Omega = \Omega_0 \cup \Omega_1 \cup \cdots \cup \Omega_n$, where $|\Omega_i| = a_i$ for $0 \leq i \leq n$. Suppose that we have an algorithm which for any inverse temperature $\beta \geq 0$ generates a random configuration from the distribution $\mu_\beta$ over $\Omega$ where the probability of a configuration $\sigma \in \Omega$ is

$$\mu_\beta(\sigma) = \frac{\mathrm{e}^{-\beta H(\sigma)}}{Z(\beta)}, \qquad (1)$$

where $H(\sigma)$ is the Hamiltonian of the configuration defined as

$$H(\sigma) = i \quad \text{such that } \sigma \in \Omega_i.$$

We now describe the details of the standard approach for using such a sampling algorithm to approximately evaluate $Z(\beta)$. In the general setting of Definition 1.1, for $X \sim \mu_\beta$, the random variable

$$W_{\beta,\beta'} := \mathrm{e}^{(\beta-\beta')H(X)} \qquad (2)$$

is an unbiased estimator for $Z(\beta')/Z(\beta)$. Indeed,

$$\mathrm{E}\left(W_{\beta,\beta'}\right) = \frac{1}{Z(\beta)} \sum_{\sigma \in \Omega} \mathrm{e}^{-\beta H(\sigma)} \cdot \mathrm{e}^{(\beta-\beta')H(\sigma)} = \frac{Z(\beta')}{Z(\beta)}. \qquad (3)$$

Thus, $a_0 = Z(\infty)$ can be approximated as follows. Take $\beta_0 < \beta_1 < \cdots < \beta_\ell$ with $\beta_0 = 0$ and $\beta_\ell = \infty$. Express $Z(\infty)$ as a telescoping product

$$Z(\infty) = Z(0)\frac{Z(\beta_1)}{Z(\beta_0)}\frac{Z(\beta_2)}{Z(\beta_1)} \cdots \frac{Z(\beta_\ell)}{Z(\beta_{\ell-1})}, \qquad (4)$$

and approximate each fraction in the product using the unbiased estimator $W_{\beta_i,\beta_{i+1}}$. The initial term $Z(0)$ is typically trivial to compute.

Taking sufficiently many samples for each $W_{\beta_i,\beta_{i+1}}$ will give a good approximation of $a_0$. The question we study in this paper is: how should one choose the inverse temperatures $\beta_0, \ldots, \beta_\ell$ so as to minimize the number of samples needed to estimate (4)? A specific choice of $\beta_0, \ldots, \beta_\ell$ is called a *cooling schedule*.

In the past, MCMC algorithms have used cooling schedules that ensure that each ratio in the telescoping product is bounded by a constant. Dyer and Frieze [2] used a non-trivial application of Chebyshev's inequality to show that $O(\ell)$ samples per ratio are sufficient to obtain an $(1 \pm \varepsilon)$ approximation of $Z(\infty)$.

For applications such as colorings or Ising model, requiring that each ratio is at most a constant, implies that the length of the cooling schedule is at least $\Omega(n)$, since $Z(0)$ and $Z(\infty)$ typically differ by an exponential factor. A general cooling schedule of length $O^*(n)$ was presented in Bezáková et al [1]. All schedules prior to our work use non-adaptive cooling schedules. By non-adaptive we refer to a schedule that depends only on $n$ and $A$ but not the structure of $Z$.

The recent volume algorithm of [11, 12] uses a non-adaptive cooling schedule of length $O(\sqrt{n})$ to estimate the volume of a convex body in $\mathbb{R}^n$. Their result relies on the logconcavity of the function $\beta^n Z(\beta)$ where $Z$ is the analogue of the partition function.

Here we present a cooling schedule for discrete partition functions with length roughly $\sqrt{\ln A}$ where $A = Z(0)$. (Note, $\sqrt{\ln A}$ is roughly $\sqrt{n}$ in the examples we have been considering here). The discrete setting presents the following new challenge. As we show in this paper, there can be no short *non-adaptive* cooling schedule for discrete partition functions. Any such non-adaptive schedule has length $\Omega(\ln A)$ in the worst case. (We defer precise statements of our results until we formally present the background material.)

Our main result is that every partition function does have an *adaptive* schedule of length roughly $\sqrt{\ln A}$. Further, the schedule can be figured out efficiently on the fly.

The existence of a short schedule follows from an interesting geometric fact: any convex function $f$ can be approximated by a piecewise linear function $g$ consisting of few pieces, see Figure 1 in Section 4 for an illustration. More precisely, $f$ is approximated in the following sense: for all $x \geq 0$, we have $0 \leq g(x) - f(x) \leq 1$.

For well-known problems such as counting colorings or matchings, and estimating the partition function of the Ising model, our results imply an improvement in the running time by a factor of $n$, since the complexity grows with the square of the schedule length; see Section 8 for a precise statement of the applications of our results.

We observe (in Section 4.1) that our techniques apply to the continuous setting as well, specifically, to the integration of general functions in $\mathbb{R}^n$. The key property required for the existence of an adaptive schedule is the *logconvexity* of the partition function $Z(\beta)$. However, this does not immediately lead to any new algorithms for integration since logconcave functions are the most general class of continuous functions for which we have efficient sampling algorithms.

In Section 2 we formalize the setup described in this introduction. The lower bound for non-adaptive schedules is formally stated as Lemma 3.3 in Section 3. The existence of a short cooling schedule is proved in Section 4, and formally stated in Theorem 4.1. The algorithm for constructing a short cooling schedule is presented in Section 5. Finally, in Section 8 we present applications of our improved cooling schedule.

## 2 Chebyshev cooling schedules

Let $W := W_{\beta,\beta'}$ be the estimator defined by (2) whose expectation is a individual ratio in the telescoping product. As usual, we will use the squared coefficient of variance $\text{Var}(W)/\text{E}(W)^2$ as a measure of the quality of the estimator $W$, namely to derive a bound on the number of samples needed for reliable estimation of $\text{E}(W)$. We will also use the quantity $\text{E}(W^2)/\text{E}(W)^2 = 1 + \text{Var}(W)/\text{E}(W^2)$.

**Lemma 2.1** (Chebyshev). *Let $W$ be a random variable with $\text{E}(W) < \infty$ and $\text{E}(W^2) < \infty$. Let $\varepsilon > 0$. We have*

$$P\big((1-\varepsilon)\text{E}(W) \leq W \leq (1+\varepsilon)\text{E}(W)\big) \geq 1 - \frac{\text{Var}(W)}{\varepsilon^2 \text{E}(W)^2} \geq 1 - \frac{\text{E}(W^2)}{\varepsilon^2 \text{E}(W)^2}.$$

The following lemma of Dyer and Frieze [2] is now well-known.

**Theorem 2.2.** *Let $W_1, \ldots, W_\ell$ be independent random variables with $\text{E}(W_i^2)/\text{E}(W_i)^2 \leq B$ for $i \in [\ell]$. Let $\widehat{W} = W_1 \ldots W_\ell$. Let $S_i$ be the average of $16B\ell/\varepsilon^2$ independent random samples from $W_i$ for $i \in [\ell]$. Let $\widehat{S} = S_1 S_2 \cdots S_\ell$. Then*

$$\Pr\left((1-\varepsilon)\text{E}\left(\widehat{W}\right) \leq \widehat{S} \leq (1+\varepsilon)\text{E}\left(\widehat{W}\right)\right) \geq 3/4.$$

It will be convenient to rewrite $\mathrm{E}(W^2)/\mathrm{E}(W)^2$ for $W := W_{\beta,\beta'}$ in terms of the partition function $Z$. We have

$$\mathrm{E}(W^2) = \frac{1}{Z(\beta)} \sum_{\sigma \in \Omega} e^{-\beta H(\sigma)} e^{2(\beta-\beta')H(\sigma)} = \frac{Z(2\beta'-\beta)}{Z(\beta)},$$

and hence

$$\frac{\mathrm{E}(W^2)}{\mathrm{E}(W)^2} = \frac{Z(2\beta'-\beta)Z(\beta)}{Z(\beta')^2}. \tag{5}$$

Equation (5) motivates the following definition.

**Definition 2.3.** Let $B > 0$ be a constant. Let $Z$ be a partition function. Let $\beta_0, \ldots, \beta_\ell$ be a sequence of inverse temperatures such that $0 = \beta_0 < \beta_1 < \cdots < \beta_\ell = \infty$. The sequence is called a *B-Chebyshev cooling schedule* for $Z$ if

$$\frac{Z(2\beta_{i+1} - \beta_i)Z(\beta_i)}{Z(\beta_{i+1})^2} \leq B, \tag{6}$$

for all $i = 0, \ldots, \ell - 1$.

The following bound on the number of samples is an immediate consequence of Theorem 2.2.

**Corollary 2.4.** *Let $Z$ be a partition function. Suppose that we are given a $B$-Chebyshev cooling schedule $\beta_0, \ldots, \beta_\ell$ for $Z$. Then, using $16B\ell^2/\varepsilon^2$ samples in total, we can compute $\widehat{S}$ such that*

$$P\big((1-\varepsilon)Z(\infty) \leq \widehat{S} \leq (1+\varepsilon)Z(\infty)\big) \geq 3/4.$$

## 3 Lower bound for non-adaptive schedules

A cooling schedule will be called *non-adaptive* if it depends only on $n$ and $A = Z(0)$ and assumes $Z(\infty) \geq 1$. Thus, such a schedule does not depend on the structure of the partition function.

The advantage of non-adaptive cooling schedules is that they do not need to be figured out on the fly. An example of a non-adaptive Chebyshev cooling schedule that works for any partition function of degree $n$, where $Z(0) = A$, is

$$0, \frac{1}{n}, \frac{2}{n}, \ldots, \frac{n \ln A}{n}, \infty. \tag{7}$$

The idea behind the schedule (7) is that small changes in the inverse temperature result in small changes of the partition function. We will state this observation more precisely, since we will use it later.

**Lemma 3.1.** *Let $\varepsilon > 0$ and let $\beta \leq \beta' \leq \beta + \varepsilon$. Let $Z$ be a partition function of degree $n$. Then*

$$Z(\beta)e^{-\varepsilon n} \leq Z(\beta') \leq Z(\beta). \tag{8}$$

**Proof :**
For $i \leq n$ we have
$$e^{-\beta i} e^{-\varepsilon n} \leq e^{-(\beta+\varepsilon)i} \leq e^{-\beta' i} \leq e^{-\beta i}. \tag{9}$$

Equation (8) now follows by applying (9) to each term of the $Z$'s in (8). ∎

To see that (7) is a Chebyshev cooling schedule, note that, by Lemma 3.1, the random variable $W_{\beta,\beta'}$ defined by (2) has values from the interval $[1/\mathrm{e}, 1]$ if $0 \leq \beta' - \beta \leq 1/n$. This implies that for $W := W_{\beta,\beta'}$ the left-hand side of (5) is bounded by a constant if $\beta, \beta' < \infty$ are neighbors in (7). It remains to show that (5) is bounded for $\beta = \ln A$ and $\beta' = \infty$. Note that that $Z(\infty) \geq 1$ (since $a_0 \geq 1$) and

$$Z(\ln A) = a_0 + \sum_{i=1}^{n} a_i e^{-i \ln A} \leq Z(\infty) + \frac{1}{A} \sum_{i=1}^{n} a_i \leq Z(\infty) + 1.$$

and hence for the right-hand side of (5) we obtain

$$\frac{Z(\ln A)}{Z(\infty)} \leq 2. \tag{10}$$

The length of the schedule (7) is $O(n \ln A)$. The following more efficient non-adaptive Chebyshev cooling schedule of length $O((\ln A) \ln n)$ is given in [1]:

$$0, \frac{1}{n}, \frac{2}{n}, \ldots, \frac{k}{n}, \frac{k\gamma}{n}, \frac{k\gamma^2}{n}, \ldots, \frac{k\gamma^t}{n}, \infty, \tag{11}$$

where $k = \lceil \ln A \rceil$, $\gamma = 1 + \frac{1}{\ln A}$, and $t = \lceil (1 + \ln A) \ln n \rceil$. The schedule (11) is based on the following observation (the statement of Lemma 3.2 slightly differs from [1] and hence we include a short proof).

**Lemma 3.2** ([1]). *Let $Z$ be a partition function with $Z(0) = A$. Let $\beta > 0$ be an inverse temperature and let $\beta' = \beta(1 + \frac{1}{\ln A})$. Then*

$$\frac{1}{2\mathrm{e}} Z(\beta) \leq Z(\beta').$$

**Proof :**
Let $n$ be the degree of $Z$. First assume that $a_n e^{-\beta n} \geq 1$. We have $a_n \leq Z(0) = A$ and hence $\beta \leq \frac{\ln A}{n}$. This implies $\beta' \leq \beta + \frac{1}{n}$ and we can use Lemma 3.1.

Now assume $a_n e^{-\beta n} < 1$. Let $k \in \{0, \ldots, n\}$ be the smallest such that

$$\sum_{i=k}^{n} a_i e^{-\beta i} < 1. \tag{12}$$

Note that $k \geq 1$, since $a_0 \geq 1$. From the minimality of $k$ we obtain

$$A e^{-\beta(k-1)} \geq \sum_{i=k-1}^{n} a_i e^{-\beta i} \geq 1,$$

and hence $\beta(k-1) \leq \ln A$. Hence for $i \leq k-1$ we have $\beta' i \leq \beta i + 1$. Now

$$Z(\beta) < 1 + \sum_{i=0}^{k-1} a_i e^{-\beta i}, \tag{13}$$

and

$$Z(\beta') \geq \sum_{i=0}^{k-1} a_i e^{-\beta' i} \geq \sum_{i=0}^{k-1} a_i e^{-\beta i - 1} \geq \frac{1}{e} \sum_{i=0}^{k-1} a_i e^{-\beta i} \geq \frac{1}{e}. \tag{14}$$

Combining (13) and (14) we obtain the result. ∎

Next we show that the schedule (11) is the best possible up to a constant factor. We will see later that *adaptive* cooling schedules can be much shorter.

**Lemma 3.3.** *Let $n \in \mathbb{Z}^+$, and $A, B \in \mathbb{R}^+$. Let $S = \beta_0, \beta_1, \ldots, \beta_\ell$ be a non-adaptive B-Chebyshev cooling schedule which works for all partition functions of degree at most $n$ with $Z(0) = A$, and $Z(\infty) \geq 1$. Assume $\beta_0 = 0$ and $\beta_\ell = \infty$. Then*

$$\ell \geq \ln(n/e) \left( \frac{\ln(A-1)}{\ln(4B)} - 1 \right). \tag{15}$$

In the proof of Lemma 3.3 we will need the following bound on the first step of the cooling schedule.

**Lemma 3.4.** *Assume $A - 1 > 4B$. Then*

$$\beta_1 \leq \frac{\ln(4B)}{n}. \tag{16}$$

**Proof of Lemma 3.4:**
Let $0 \leq a \leq A - 1$. Then $S$ has to be a $B$-Chebyshev cooling schedule for

$$Z(\beta) = \frac{A}{1+a} \left( 1 + a e^{-\beta n} \right).$$

The equation (6) needs to be satisfied for $Z$, $\beta_0 = 0$ and $\beta_1$. Thus

$$\frac{(1 + a e^{-2\beta_1 n})(1 + a)}{(1 + a e^{-\beta_1 n})^2} \leq B. \tag{17}$$

After substitution $z = e^{-\beta_1 n}$, equation (17) becomes equivalent to

$$\frac{(1 + a z^2)(1 + a)}{(1 + a z)^2} = 1 + a \left( \frac{1-z}{1+az} \right)^2 \leq B. \tag{18}$$

Suppose that $z \leq \frac{1}{A-1}$. Note that the left-hand side of (18) is decreasing in $z$. Hence, (18) is true for $z = \frac{1}{A-1}$. Let $a = A - 1$. For this choice of $a$ and $z$, (18) yields $(A-1)/4 \leq B$, a contradiction with $A > 4B + 1$. Thus, $z > \frac{1}{A-1}$.

Since $z > 1/(A-1)$, we have $1/z < A - 1$ and, hence, we can choose $a = 1/z$. Plugging $a = 1/z$ into (18) we obtain

$$\frac{(1+z)^2}{4z} \leq B, \tag{19}$$

and, hence, $z \geq 1/(4B)$, which implies (16). ∎

The Lemma 3.4 immediately gives a bound on the later steps in the schedule. If the current inverse temperature is $\beta_i$ then the coefficient of degree $k$ is decimated by $\mathrm{e}^{-\beta_i k}$.

**Corollary 3.5.** *Let $k \in \{1, \ldots, n\}$. Assume $(A-1)\mathrm{e}^{-\beta_i k} > 4B$. Then*

$$\beta_{i+1} - \beta_i \leq \frac{\ln(4B)}{k}.$$

**Proof of Lemma 3.3:**
Let $S' = \beta_0, \beta_1, \ldots, \beta_\ell$ be the shortest sequence such that $\beta_0 = 0, \beta_\ell = \infty$ and the Corollary 3.5 is satisfied for $S'$.

We can greedily construct the shortest sequence $S'$ as follows. If $k \in \{1, \ldots, n\}$ is the largest such that $(A-1)\mathrm{e}^{-\beta_i k} > 4B$ then we take

$$\beta_{i+1} = \beta_i + \frac{\ln(4B)}{k}.$$

(If $(A-1)\mathrm{e}^{-\beta_i} \leq 4B$ then we take $\beta_{i+1} = \infty$.)

Let $x_i$ be the number of indices for which $\beta_{i+1} - \beta_i = \frac{\ln(4B)}{i}$. Let $j \in \{2, \ldots, n\}$ and

$$\beta = \sum_{i=j}^{n} x_i \frac{\ln(4B)}{i}. \tag{20}$$

From $\beta$ we take a step of length at least $\frac{\ln(4B)}{j-1}$ (since we already took all shorter steps) and hence

$$(A-1)\mathrm{e}^{-\beta j} \leq 4B. \tag{21}$$

Plugging (20) into (21) we obtain

$$\sum_{i=j}^{n} x_i \frac{\ln(4B)}{i} \geq \frac{1}{j} \ln \frac{A-1}{4B}. \tag{22}$$

Summing (22) for $j = 2, \ldots, n$ we obtain

$$(\ln(4B)) \sum_{j=2}^{n} x_i \geq \left( \sum_{j=2}^{n} \frac{1}{j} \right) \ln \frac{A-1}{4B} \geq \left( \ln \frac{n}{\mathrm{e}} \right) \ln \frac{A-1}{4B},$$

which implies (15). ∎

The number of samples needed in Theorem 2.2 (and Corollary 2.4) is linear in $B$ and hence, in view of Lemma 3.3, the optimal value of $B$ is a constant. Our understanding of non-adaptive schedules is now complete up to a constant factor. In particular, the schedule (11) and Lemma 3.3 imply that the optimal non-adaptive schedule has length $\Theta((\ln A) \ln n)$.

We would like to have a similar understanding of adaptive cooling schedules. A reasonable conjecture is that the optimal adaptive schedule has length

$$\Theta\left(\sqrt{(\ln A) \ln n}\right). \tag{23}$$

We will present an adaptive schedule of length $O\left(\sqrt{\ln A}(\ln n)\ln\ln A\right)$. This comes reasonably close to our guess in (23) (in fact, in our applications we are only off by polylogarithmic factors).

We will have the following technical assumptions on $A$ and $n$.

$$\ln n \geq 1, \quad \ln\ln A \geq 1, \text{ and } \quad A \geq \ln n. \tag{24}$$

The first two assumptions are necessary since both $\ln n$ and $\ln\ln A$ figure in our bounds on the length of the schedule. The third assumption is justified for the following two reasons. First, in the applications we consider, $A$ is usually exponential in $n$. Second, if $A$ is too small then no cooling schedule is necessary - a direct application of the Monte Carlo method uses only $A/\varepsilon^2$ samples (which, for $A \leq \ln n$, is less than the number of samples needed by a cooling schedule of length given by (23)).

## 4   Adaptive cooling schedules

In this section, we prove the existence of short adaptive cooling schedules for general partition functions. We now formally state the result (to simplify the exposition we will choose $B = \mathrm{e}^2$, the construction works for any $B$).

**Theorem 4.1.** *Let $Z$ be a partition function of degree $n$. Let $A = Z(0)$. Assume that $Z(\infty) \geq 1$. There exists an $\mathrm{e}^2$-Chebyshev cooling schedule $S$ for $Z$ whose length is at most*

$$4(\ln\ln A)\sqrt{(\ln A)\ln n}.$$

It will be convenient to define $f(\beta) = \ln Z(\beta)$. Some useful properties of $f$ are summarized in the next lemma. We include a short proof in Section 6.

**Lemma 4.2.** *Let $f(\beta) = \ln Z(\beta)$ where $Z$ is a partition function of degree $n$. Then (a) $f$ is decreasing, (b) $f'$ is increasing (i. e., $f$ is convex) (c) $f'(0) \geq -n$.*

Recall that an $\mathrm{e}^2$-Chebyshev cooling schedule for $Z$ is a sequence of inverse temperatures $\beta_0, \beta_1, \ldots, \beta_\ell$ such that $\beta_0 = 0$, $\beta_\ell = \infty$, and

$$\frac{Z(2\beta_{i+1} - \beta_i)Z(\beta_i)}{Z(\beta_{i+1})^2} \leq \mathrm{e}^2. \tag{25}$$

Since (25) is invariant under scaling we can, without loss of generality, assume $Z(\infty) = 1$ (or equivalently $a_0 = 1$). Since we assumed $a_0 \geq 1$ the scaling will not increase $Z(0)$.

Let $f(\beta) = \ln Z(\beta)$, so that $f(0) = \ln A$, and $f(\infty) = 0$. The condition (25) is equivalent to

$$\frac{f(2\beta_{i+1} - \beta_i) + f(\beta_i)}{2} - f(\beta_{i+1}) \leq 1. \tag{26}$$

If we substitute $x = \beta_i$ and $y = 2\beta_{i+1} - \beta_i$, the condition can be rewritten as

$$f\left(\frac{x+y}{2}\right) \geq \frac{f(x) + f(y)}{2} - 1.$$

In words, $f$ satisfies approximate concavity. The main idea of the proof is that we do not require this property to hold everywhere but only in a sparse subset of points which will correspond to the cooling schedule. A similar viewpoint is that we will show that $f$ can be approximated by a piecewise linear function $g$ with few pieces, see Figure 1 for an illustration. We form the segments of $g$ in the following inductive, greedy manner. Let $\gamma_i$ denote the endpoint of the last segment. We then set $\gamma_{i+1}$ as the maximum value such that the midpoint $m_i$ of the segment $(\gamma_i, \gamma_{i+1})$ satisfies (26) (for $\beta_i = \gamma_i, \beta_{i+1} = m_i$). We now formally state the lemma on the approximation of $f$ by a piecewise linear function.

**Lemma 4.3.** *Let $f : [0, \gamma] \mapsto \mathbb{R}$ be a decreasing, convex function. There exists a sequence $\gamma_0 = 0 < \gamma_1 < \cdots < \gamma_j = \gamma$ such that for all $i \in \{0, \ldots, j-1\}$,*

$$f\left(\frac{\gamma_i + \gamma_{i+1}}{2}\right) \geq \frac{f(\gamma_i) + f(\gamma_{i+1})}{2} - 1, \tag{27}$$

*and*

$$j \leq 1 + \sqrt{(f(0) - f(\gamma)) \ln \frac{f'(0)}{f'(\gamma)}}.$$

**Proof :**
Let $\gamma_0 := 0$. Suppose that we already constructed the sequence up to $\gamma_i$. Let $\gamma_{i+1}$ be the largest number from the interval $[\gamma_i, \gamma]$ such that (27) is satisfied. Let $m_i = (\gamma_i + \gamma_{i+1})/2$, let $\Delta_i = (\gamma_{i+1} - \gamma_i)/2$, and $K_i = f(\gamma_i) - f(\gamma_{i+1})$.

If $\gamma_{i+1} = \gamma$ then we are done constructing the sequence. Otherwise, by the maximality of $\gamma_{i+1}$, we have

$$f(m_i) = \frac{f(\gamma_i) + f(\gamma_{i+1})}{2} - 1. \tag{28}$$

Using the convexity of $f$ and (28) we obtain

$$-f'(\gamma_i) \geq \frac{f(\gamma_i) - f(m_i)}{\Delta} = \frac{K_i + 2}{2\Delta}, \tag{29}$$

and

$$-f'(\gamma_{i+1}) \leq \frac{f(f(m_i) - \gamma_{i+1})}{\Delta} = \frac{K_i - 2}{2\Delta}. \tag{30}$$

Combining (29) and (30) we obtain

$$\frac{f'(\gamma_{i+1})}{f'(\gamma_i)} = \frac{-f'(\gamma_{i+1})}{-f'(\gamma_i)} \leq \frac{K_i - 2}{K_i + 2} = 1 - \frac{4}{K_i + 2}. \tag{31}$$

From (30) and the fact that $f$ is decreasing we obtain $K_i \geq 2$. Hence we can estimate (31) as follows

$$\frac{f'(\gamma_{i+1})}{f'(\gamma_i)} \leq 1 - \frac{4}{K_i + 2} \leq 1 - \frac{1}{K_i} \leq e^{-1/K_i}. \tag{32}$$

Since $f$ is decreasing, we have

$$\sum_{i=0}^{j-2} K_i \leq f(0) - f(\gamma). \tag{33}$$

Now we combine (32) for all $i \in \{0, \ldots, j-2\}$.

$$\sum_{i=0}^{j-2} \frac{1}{K_i} \leq \ln \frac{f'(0)}{f'(\gamma)}. \tag{34}$$

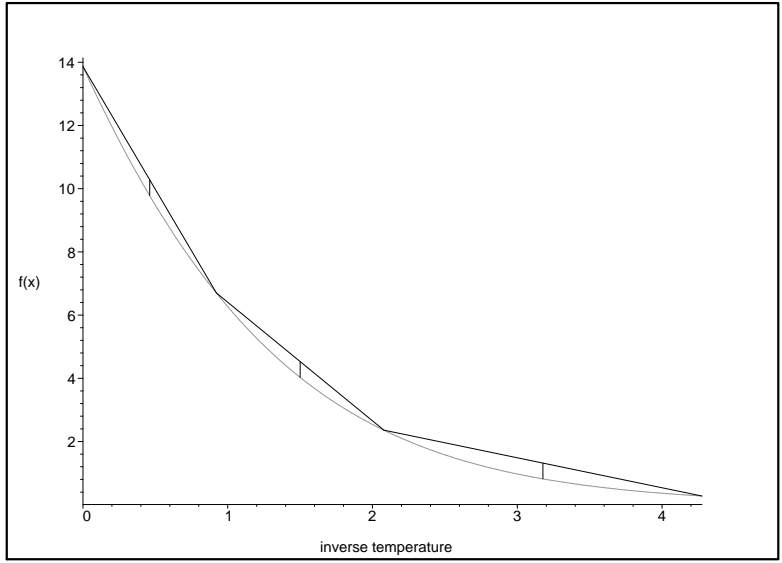

Figure 1: The light curve is $f(x) = \ln Z(x)$ for the partition function $Z(x) = (1 + \exp(-x))^{20}$. The dark curve is a piecewise linear function $g$ consisting of 3 pieces which approximates $f$. In particular, $g \geq f$ and the midpoint of each piece is close to the average of the endpoints (specifically, (25) holds).

Applying Cauchy-Schwarz inequality on (33) and (34) we obtain

$$(j-1)^2 \leq (f(0) - f(\gamma)) \ln \frac{f'(0)}{f'(\gamma)}.$$

∎

The construction immediately yields a natural cooling schedule. A schedule ending at $\beta_k = \gamma_i$, can now be extended by $\beta_{k+1} = m_i$ where $m_i$ is the midpoint of the segment $(\gamma_i, \gamma_{i+1})$. Moreover, we can then set $\beta_{k+2}$ as the midpoint of $(m_i, \gamma_{i+1})$. We continue in this geometric manner for at most $\ln \ln A$ steps, after which we can set the next inverse temperature in our schedule to $\gamma_{i+1}$. Then we continue on the next segment. It then follows that the length $\ell$ of the cooling schedule satisfies $\ell \leq j \ln \ln A$ where $j$ is the length of the sequence from Lemma 4.3. We now present the proof of the Theorem 4.1.

**Proof of Theorem 4.1:**

Let $\gamma$ be such that $f(\gamma) = 1$. We describe a sequence $\beta_0 = 0 < \beta_1 < \ldots \beta_\ell = \gamma$ satisfying (26). Note that since $f(\gamma) = 1$, we can take $\beta_{\ell+1} = \infty$ and the sequence will still satisfy (26) (and thus we get a complete $e^2$-Chebyshev cooling schedule for $Z$). We have

$$Z(\gamma) = \exp(f(\gamma)) = \sum_{i=0}^{n} a_i e^{-i\gamma} = e,$$

and, hence, (using $a_0 = 1$)

$$-Z'(\gamma) = \sum_{i=0}^{n} i a_i e^{-i\gamma} \geq e - 1.$$

Thus
$$-f'(\gamma) = -\ln Z(\gamma) = \frac{-Z'(\gamma)}{Z(\gamma)} = \frac{\sum_{i=0}^n i a_i e^{-i\gamma}}{\sum_{i=0}^n a_i e^{-i\gamma}} \geq \frac{e-1}{e}. \tag{35}$$

By Lemma 4.3, there exists a sequence of $\gamma_0 = 0 < \gamma_1 < \cdots < \gamma_j = \gamma$ of length
$$j \leq 1 + \sqrt{(\ln A) \ln \frac{n e}{e - 1}} \tag{36}$$

such that (27) is satisfied.

Now we show how to add $\lceil \ln \ln A \rceil$ inverse temperatures between each pair $\gamma_i$ and $\gamma_{i+1}$ to obtain our cooling schedule. For notational convenience we show this only for $\gamma_0 = 0$ and $\gamma_1$.

Note that (27) implies that (26) is satisfied for $\beta_0 = 0$ and $\beta_1 = \gamma_1/2$. We now show that
$$0, (1/2)\gamma_1, (3/4)\gamma_1, (7/8)\gamma_1, \ldots, (1 - 2^{-\lceil \ln \ln A \rceil})\gamma_1, \gamma_1$$

is an $e^2$-Chebyshev cooling schedule. Let
$$g(x) = f\left(\frac{\gamma_1 + x}{2}\right) - \frac{f(x) + f(\gamma_1)}{2}.$$

Note that by (28) we have $g(0) = -1$. We have
$$g'(x) = \frac{1}{2}\left(f'\left(\frac{\gamma_1 + x}{2}\right) - f'(x)\right).$$

Thus
$$\text{if } x \leq \gamma_1 \text{ we have } g'(x) \geq 0, \tag{37}$$

and, hence,
$$g(x) \geq g(0) = -1.$$

Plugging in $x = (1 - 2^{-t})\gamma_1$ we conclude
$$f((1 - 2^{-t-1})\gamma_1) \geq \frac{f((1 - 2^{-t})\gamma_1) + f(\gamma_1)}{2} - 1. \tag{38}$$

From (28) and (38) it follows that the sequence
$$0, (1/2)\gamma_1, (3/4)\gamma_1, (7/8)\gamma_1, \ldots, (1 - 2^t)\gamma_1, \gamma_1 \tag{39}$$

satisfies (26). We will now show that we can truncate the sequence at $t = \lceil \ln \ln A \rceil$ and take the next step to $\gamma_1$.

By the convexity of $f$
$$f((1 - 2^{-t-1})\gamma_1) \leq \frac{f((2 - 2^{-t})\gamma_1) + f(\gamma_1)}{2},$$

and hence
$$f((1 - 2^{-t-1})\gamma_1) - f(\gamma_1) \leq \frac{f((1 - 2^{-t})\gamma_1) - f(\gamma_1)}{2}. \tag{40}$$

The equation (40) states that the distance of $f((1-2^{-t})\gamma_1)$ from $f(\gamma_1)$ halves in each step. Recall that $f(\gamma_1/2) - f(\gamma_1) \leq f(0) \leq \ln A$ and, hence, for $t := \lceil \ln \ln A \rceil$ we have

$$f((1-2^{-t})\gamma_1) - f(\gamma_1) \leq 1. \tag{41}$$

This completes the construction of the cooling schedule. The length of the schedule is $\leq jt$. Plugging in (36) yields the theorem.

∎

The optimal Chebyshev cooling schedule can be obtained in a greedy manner. In particular, starting with $\beta_0 = 0$, and then from $\beta_i$, choosing the maximum $\beta_{i+1}$ for which (26) is satisfied. The reason why the greedy strategy works is that if we can step from $\beta$ to $\beta'$, then for any $\gamma \in [\beta, \beta']$ we can step from $\gamma$ to $\beta'$ (i.e., having large inverse temperature can not hurt us). The last fact follows from the convexity of $f$ (or alternatively from (37)).

**Corollary 4.4.** *Let $Z$ be a partition function of degree $n$. Let $A = Z(0)$. Assume that $Z(\infty) \geq 1$. Suppose that $\beta_0 < \cdots < \beta_\ell$ is a cooling schedule for $Z$. Then the number of indices $i$ for which*

$$\frac{Z(2\beta_{i+1} - \beta_i)Z(\beta_i)}{Z(\beta_{i+1})^2} \geq e^2 \tag{42}$$

*is at most $4(\ln \ln A)\sqrt{(\ln A) \ln n}$.*

We now formally prove the greedy property of Chebyshev cooling schedules. Note that we can make a step from $x$ to $y$ if $g(x, y) \leq 1$, where

$$g(x, y) = \frac{f(x) + f(2y - x)}{2} - f(y). \tag{43}$$

**Lemma 4.5.** *Let $Z$ be a partition function. Let $f = \ln Z(\beta)$ and let $g$ be given by (43). The function $g(x, y)$ is decreasing in $x$ for $x < y$. The function $g(x, y)$ is increasing in $y$ for $x < y$.*

**Proof :**
By Lemma 4.2 we have that $f'$ is an increasing function. We have $2y - x > x$ and hence

$$\frac{\partial g(x, y)}{\partial x} = \frac{f'(x) - f'(2y - x)}{2} < 0.$$

Analogously

$$\frac{\partial g(x, y)}{\partial y} = f'(2y - x) - f'(y) > 0.$$

∎

**Proof of Corollary 4.4:**
Let $k_0 < k_1 < \cdots < k_m$ be the indices for which (42) is satisfied. Let $\alpha_0 = 0 < \alpha_1 < \cdots < \alpha_\ell = \infty$ be the optimal $e^2$-Chebyshev cooling schedule. We are going to show, using induction on $j$, that

$$\alpha_j \leq \beta_{k_j}. \tag{44}$$

Clearly (44) is true for $j = 0$.

Now assume (44) is true for some $j$. We have $g(\alpha_j, \alpha_{j+1}) \leq 1$, $g(\beta_i, \beta_{i+1}) \geq 1$, and $\alpha_j \leq \beta_i$. From Lemma 4.5 it follows that $\alpha_{j+1} \leq \beta_{i+1}$ and hence

$$\alpha_{j+1} \leq \beta_{i+1} \leq \beta_{k_{j+1}},$$

completing the induction step.

Equation (44) implies $m \leq \ell \leq 4(\ln \ln A)\sqrt{(\ln A) \ln n}$. ∎

## 4.1 Extensions

The key property of $Z(\beta)$ used in the proof of existence of a fast cooling schedule is the fact that it is logconvex (i.e., its logarithm, $f(\beta) = \ln Z(\beta)$, is convex). The proof above can be appropriately modified for other function classes with this property. We highlight this for a class of continuous functions.

**Lemma 4.6.** *Let $g : \mathbb{R}^n \to \mathbb{R}$ be a continuous, integrable, nonnegative function. Define*

$$Z(\beta) = \int_{\mathbb{R}^n} g(x)^\beta \, dx$$

*for $\beta > 0$. Then $Z(\beta)$ is logconvex.*

The proof is identical to that of Lemma 4.2, part(b).

## 4.2 Lower bound for adaptive cooling

**Lemma 4.7.** *Let $n \geq 1$. Consider the following partition function of degree $n$:*

$$Z(\beta) = (1 + \mathrm{e}^{-\beta})^n.$$

*Any B-Chebyshev cooling schedule for $Z(\beta)$ has length at least $\sqrt{n/(20 \ln B)}$.*

**Proof :**
Let $f(\beta) = \ln Z(\beta) = n \ln(1 + \mathrm{e}^{-\beta})$. If the current inverse temperature is $\beta_i =: \beta$, the next inverse temperature $\beta_{i+1} =: \beta + x$ has to satisfy

$$f(\beta) + f(\beta + 2x) - 2f(\beta + x) \leq \ln B.$$

Later we will show that for any $\beta \in [0,1]$ and $x \in [0,1]$ we have

$$f(\beta) + f(\beta + 2x) - 2f(\beta + x) \geq \frac{n}{20} x^2. \tag{45}$$

From (45) it follows that for $\beta \leq 1$ the inverse temperature increases by at most

$$x \leq \sqrt{\frac{20 \ln B}{n}},$$

and, hence, the length of the schedule is at least $\sqrt{n/(20 \ln B)}$.

It remains to show (45). Let

$$g(x, \beta) := \frac{f(\beta) + f(\beta + 2x) - 2f(\beta + x)}{2n}.$$

We have
$$\frac{\partial}{\partial x}g(x,\beta) = \frac{e^{-\beta-x}}{1+e^{-\beta-x}} - \frac{e^{-\beta-2x}}{1+e^{-\beta-2x}}.$$
We will show
$$\frac{e^{-\beta-x}}{1+e^{-\beta-x}} - \frac{e^{-\beta-2x}}{1+e^{-\beta-2x}} \geq x/20, \qquad (46)$$
which will imply (45) (by integration over $x$).

Let $C := e^{-\beta}$ and $y := 1 - e^{-x}$. Note that $C \in [1/e, 1]$, $y \in [0, 1-1/e]$, and $x = -\ln(1-y)$. For $y \in [0, 1-1/e]$ we have $-\ln(1-y) \leq y + y^2$ and hence it is enough to show
$$\frac{C(1-y)}{1+C(1-y)} - \frac{C(1-y)^2}{1+C(1-y)^2} \geq \frac{1}{20}(y+y^2). \qquad (47)$$
Multiplying both sides by the numerators we obtain that (47) is equivalent to
$$P(y,C) := y(y+1)(y-1)^3 C^2 - (y^4 - 2y^3 + 19y^2 - 18y)C - (y^2+y) \geq 0.$$
The polynomial $y(y+1)(y-1)^3$ is negative for our range of $y$ and hence for any fixed $y$, the minimum of $P(y,C)$ over $C \in [1/3, 1]$ occurs either at $C = 1$ or at $C = 1/3$ (we only need to show positivity of $P(y,C)$ for $C \in [1/e, 1]$, but for numerical convenience we show it for a larger interval). We have
$$p(y,1) = y^5 - 3y^4 + 2y^3 - 18y^2 + 16y, \qquad (48)$$
and
$$9p(y,1/3) = y^5 - 5y^4 + 6y^3 - 64y^2 + 44y. \qquad (49)$$
Both (48) and (49) are non-negative for our range of $y$ (as is readily seen by the method of Sturm sequences). This finishes the proof of (46), which in turn implies (45). ∎

## 5 An adaptive cooling algorithm

The main theorem of the previous section proves the existence of a short adaptive cooling schedule, whereas in Section 3 we proved any non-adaptive cooling schedule is much longer. In this section, we present an adaptive algorithm to find a short cooling schedule. We state the main result before describing the details of the algorithm. The algorithm has access to a sampling oracle, which on input $\beta$ produces a random sample from the distribution $\mu_\beta$, defined by (1) (or a distribution sufficiently close to $\mu_\beta$).

**Theorem 5.1.** *Let $Z$ be a partition function. Assume that we have access to an (approximate) sampling oracle from $\mu_\beta$ for any inverse temperature $\beta$. Let $\delta' > 0$. With probability at least $1 - \delta'$, algorithm PRINT-COOLING-SCHEDULE outputs a $B$-Chebyshev cooling schedule for $Z$ (with $B = 3 \cdot 10^6$), where the length of the schedule is at most*
$$\ell \leq 38\sqrt{\ln A}(\ln n)\ln\ln A. \qquad (50)$$
*The algorithm uses at most*
$$Q \leq 10^7 (\ln A)\bigl((\ln n) + \ln\ln A\bigr)^5 \ln\frac{1}{\delta'} \qquad (51)$$
*samples from the $\mu_\beta$-oracles. The samples output by the oracles have to be from a distribution $\mu'_\beta$ which is within variation distance $\leq \delta'/(2Q)$ from $\mu_\beta$.*

In Section 7 we extend the algorithm to the setting of warm-start sampling oracles (see Theorem (7.5)).

## 5.1 High-level Algorithm Description

We begin by presenting the high-level idea of our algorithm. Ideally we would like to find a sequence $\beta_0 = 0 < \beta_1 < \cdots < \beta_\ell = \infty$ such that, for some constants $1 < c_1 < c_2$, for all $i$, the random variable $W := W_{\beta_i,\beta_{i+1}}$ satisfies

$$c_1 \leq \frac{\mathrm{E}\left(W^2\right)}{\mathrm{E}\left(W\right)^2} \leq c_2. \tag{52}$$

The upper bound in (52) is necessary so that Chebyshev's inequality guarantees that few samples of $W$ are required to obtain a close estimate of the ratio $Z(\beta_i)/Z(\beta_{i+1})$. On the other side, the lower bound would imply that the length of the cooling schedule is close to optimal. We will guarantee the upper bound for every pair of inverse temperatures, but we will only obtain the lower bound for a sizable fraction of the pairs. Then, using Corollary 4.4, we will argue that the schedule is short.

During the course of the algorithm we will try to find the next inverse temperature $\beta_{i+1}$ so that (52) is satisfied. For this we will need to estimate $u = u(\beta_i, \beta_{i+1}) := \mathrm{E}\left(W^2\right)/\mathrm{E}\left(W\right)^2$. We already have an expression for $u$, given by equation (5):

$$u = \frac{\mathrm{E}\left(W^2\right)}{\mathrm{E}\left(W\right)^2} = \frac{Z(2\beta_{i+1} - \beta_i)Z(\beta_i)}{Z(\beta_{i+1})^2} = \frac{Z(2\beta_{i+1} - \beta_i)}{Z(\beta_{i+1})} \frac{Z(\beta_i)}{Z(\beta_{i+1})}. \tag{53}$$

Hence, to estimate $u$ it suffices to estimate the ratios $Z(2\beta_{i+1}-\beta_i)/Z(\beta_i)$ and $Z(\beta_i)/Z(\beta_{i+1})$. Recall that the goal of estimating $u$ was to show that $W$ is an efficient estimator of $Z(\beta_i)/Z(\beta_{i+1})$. Now it seems that to estimate $u$ we already need a good estimator for $W$. An important component of our algorithm, which allows us to escape from this circular loop, is a rough estimator for $u$ which bypasses $W$.

Recall, the Hamiltonian $H$ takes values in $\{0, 1, \ldots, n\}$. For the purposes of estimating $u$ it will suffice to know the Hamiltonian within some relative accuracy. Thus, we partition $\{0, 1, \ldots, n\}$ into (discrete) intervals of roughly equivalent values of the Hamiltonian. Since we need relative accuracy the size of the interval is smaller for smaller values of the Hamiltonian (specifically, value $i$ is an interval of size about $i/\sqrt{\ln A}$). We let $P$ denote the set of intervals. We will define the intervals so that the number of intervals $|P|$ is at most $O(\sqrt{\ln A} \ln n)$.

The rough estimator for $u$ needs an interval $I = [b, c] \subseteq \{1, \ldots, n\}$ which contributes a significant portion to $Z(\beta)$ for all $\beta \in [\beta_i, 2\beta_{i+1} - \beta_i]$. We say such an $I$ is *heavy* for that interval of inverse temperatures. Thus, if we generate a random sample from $\mu_\beta$ we have a significant probability that the sample is in the interval $I$. The key observation is that if an interval $I$ is heavy for inverse temperatures $\beta_1$ and $\beta_2$, then by generating samples from $\mu_{\beta_1}$ and $\mu_{\beta_2}$, and looking at the proportion of samples whose Hamiltonian falls into interval $I$, we can roughly estimate $Z(\beta_2)/Z(\beta_1)$.

Thus, if an interval $I$ is heavy for an interval of inverse temperatures $B = [\beta_i, \beta^*]$, then we can find a $\beta_{i+1} \in B' = [\beta_i, (\beta_i + \beta^*)/2]$ satisfying (52) (making an optimal move in some sense) or determine there is no such $\beta_{i+1} \in B'$.

In the later case we construct a sequence of inverse temperatures that goes from $\beta_i$ to $\beta^*$ where the upper bound in (52) holds for this sequence. We will show that $O(\ln \ln A)$ intermediate inverse temperatures are sufficient to go from $\beta_i$ to $\beta^*$ (the construction is analogous to the sequence (39) in the proof of Theorem 4.1). Once we reach $\beta^*$ we will be done with this interval $I$ and will not need to consider it again.

An important fact is that for an interval $I$, the set of $\beta$'s where $I$ is heavy is itself an interval. Hence, each interval causes a non-optimal step at most once (causing a sequence of $O(\ln \ln A)$ intermediate inverse temperatures). Thus, our algorithm will find a cooling schedule whose length is at most

$$O\left((\ln \ln A)\sqrt{(\ln A) \ln n} + \sqrt{\ln A}(\ln n) \ln \ln A\right), \tag{54}$$

where the first term comes from Theorem 4.1 and the second term comes from the upper bound on $|P| = O(\sqrt{\ln A}(\ln n))$ and the fact that the non-optimal steps cause the algorithm to output a sequence of $O(\ln \ln A)$ intermediate inverse temperatures.

To simplify the high-level exposition of the algorithm we glossed over a technical aspect of the rough estimator which sometimes does not allow a move long enough to finish off the interval $I$. Such a move will be long relative to the reciprocal of the width of the $I$ and will be referred to as "long" step. ("Long" steps will be analyzed by a separate argument, and their number will be smaller than (54).) Thus, in the detailed description of the algorithm we will have three kinds of steps: "optimal" steps, "interval" steps, and "long" steps.

Combining Theorem 5.1 with Corollary 2.4 we obtain.

**Corollary 5.2.** *Let $Z$ be a partition function. Let $\varepsilon > 0$ be the desired precision. Suppose that we are given access to oracles which sample from the distribution within variation distance*

$$\frac{\varepsilon^2}{10^8(\ln A)\big((\ln n) + \ln \ln A\big)^5}$$

*from $\mu_\beta$ for any inverse temperature $\beta$.*

*Using $\frac{10^{10}}{\varepsilon^2}(\ln A)\big((\ln n) + \ln \ln A\big)^5$ samples in total, we can obtain a random variable $\widehat{S}$ such that*

$$P\big((1-\varepsilon)Z(\infty) \le \widehat{S} \le (1+\varepsilon)Z(\infty)\big) \ge 3/4.$$

## 5.2 Detailed Algorithm Description

Here we present a detailed description of the algorithm. We also present pseudocode for the algorithm in Section 10 of the Appendix.

First, we construct a partition $P$ of $\{0, \ldots, n\}$ into $O(\sqrt{\ln A} \ln n)$ disjoint intervals. We construct $P$ inductively, starting with interval $[0, 0]$. Suppose that $\{0, \ldots, b-1\}$ is already partitioned. Let

$$w := \lfloor b/\sqrt{\ln A} \rfloor. \tag{55}$$

Add the interval $[b, b+w]$ to $P$ and continue inductively on $\{b+w+1, \ldots, n\}$. Note, the initial $\sqrt{\ln A}$ intervals are of size 1 (i.e., contain one natural number), and have width 0. Later (in Section 5.3) we will show the following explicit upper bound on the number of intervals in $P$.

**Lemma 5.3.** $|P| \leq 4\sqrt{\ln A} \ln n$.

In each stage of the algorithm we want an interval which is *heavy* in the following precise sense.

**Definition 5.4.** Let $Z$ be a partition function. Let $\beta \in \mathbb{R}^+$ be an inverse temperature. Let $I = [b,c] \subseteq \{0,\ldots,n\}$ be an interval. For $h \in (0,1)$, we say that $I$ is *h-heavy* for $\beta$, if for $X$ chosen from $\mu_\beta$, we have

$$\Pr(H(X) \in I) \geq h.$$

The following property will be crucial for our algorithm: the set of inverse temperatures for which an interval $I$ is heavy is itself an interval (in $\mathbb{R}^+$), the proof is deferred to Section 6.

**Lemma 5.5.** *Let $Z$ be a partition function. Let $I = [b,c] \subseteq \{0,\ldots,n\}$ be an interval. Let $h \in (0,1]$. The set of inverse temperatures for which $I$ is h-heavy forms an interval (possibly empty).*

Let

$$h := \frac{1}{8|P|}. \tag{56}$$

In our algorithm we will use an interval which is $h$-heavy. Given access to a sampler for $X \sim \mu_\beta$ one can approximately check whether an interval is $h$-heavy for $\beta$. More precisely, we can distinguish the case when $I$ is $h$-heavy versus when $I$ is not $4h$-heavy. We formalize this observation in Lemma 5.7. First we need the following definition.

**Definition 5.6.** Let $Z$ be a partition function. Let $I = [b,c] \subseteq \{0,\ldots,n\}$ be an interval. Let $\delta \in (0,1]$ and let $\beta$ be an inverse temperature. Let $X \sim \mu_\beta$ and let $Y$ be the indicator function for the event $H(X) \in I$. Let $s = \lceil (8/h) \ln \frac{1}{\delta} \rceil$. Let $U$ be the average of $s$ independent samples from $Y$. Let

$$\text{Is-Heavy}(I, \beta) = \begin{cases} \text{true} & \text{if } U \geq 2h \\ \text{false} & \text{if } U < 2h \end{cases}$$

**Lemma 5.7.** *If $I$ is not $h$-heavy at inverse temperature $\beta$, then*

$$\Pr(\text{Is-Heavy}(I, \beta) = \text{true}) \leq \delta. \tag{57}$$

*If $I$ is $4h$-heavy at inverse temperature $\beta$, then*

$$\Pr(\text{Is-Heavy}(I, \beta) = \text{false}) \leq \delta. \tag{58}$$

The above lemma is proved in Section 6.

If we take $s = \lceil (8/h) \ln \frac{1}{\delta} \rceil$ samples from $\mu_\beta$, and take the interval which received the most samples, then we are likely to get an $h$-heavy interval. Note that by our choice of $h$ there exists a $8h$-heavy interval $J$. By Lemma 5.7, it is very likely that $J$ receives more than $2hs$ samples and that all intervals which are not $h$-heavy receive less than $2hs$ samples. Thus, the interval with the most samples will likely be $h$-heavy.

**Corollary 5.8.** *Given an inverse temperature $\beta$, using $s = \lceil (8/h) \ln \frac{1}{\delta} \rceil$ samples from $\mu_\beta$ we can find an h-heavy interval. The failure probability of the procedure is at most $\delta |P|$.*

We will need a more general version of Corollary 5.8 in which the set of intervals that can be chosen is restricted. The forbidden intervals will not be $8h$-heavy and, hence, there will exist an allowed interval which is $8h$-heavy. Using the same reasoning as we used for Corollary 5.8 we obtain the following procedure, which we call FIND-HEAVY.

**Corollary 5.9.** *Let $\beta$ be an inverse temperature. Let Bad be a set of intervals such than no interval in Bad is $8h$-heavy at $\beta$. Given an inverse temperature $\beta$, using $s = \lceil (8/h) \ln \frac{1}{\delta} \rceil$ samples from $\mu_\beta$ we can find an h-heavy interval which is not in Bad. The failure probability of the procedure FIND-HEAVY is at most $\delta |P|$.*

We use the following idea: if a narrow interval is heavy for two nearby inverse temperatures $\beta_1, \beta_2$ then the interval can be used to estimate the ratio of $Z(\beta_1)$ and $Z(\beta_2)$.

**Lemma 5.10.** *Let $Z$ be a partition function. Let $I = [b, c] \subseteq \{0, \ldots, n\}$ be an interval. Let $\delta \in (0, 1]$. Suppose that $I$ is h-heavy for inverse temperatures $\beta_1, \beta_2 \in \mathbb{R}^+$. Assume that*

$$|\beta_1 - \beta_2| \cdot (c - b) \leq 1. \tag{59}$$

*For $k = 1, 2$ we define the following. Let $X_k \sim \mu_{\beta_k}$ and let $Y_k$ be the indicator function for the event $H(X_k) \in I$. Let $s = \lceil (8/h) \ln \frac{1}{\delta} \rceil$. Let $U_k$ be the average of $s$ independent samples from $Y_k$. Let*

$$\text{EST}(I, \beta_1, \beta_2) := \frac{U_1}{U_2} \exp(b(\beta_1 - \beta_2)). \tag{60}$$

*With probability at least $1 - 4\delta$ we have*

$$\frac{Z(\beta_2)}{4\mathrm{e} Z(\beta_1)} \leq \text{EST}(I, \beta_1, \beta_2)) \leq \frac{4\mathrm{e} Z(\beta_2)}{Z(\beta_1)}. \tag{61}$$

The above lemma is proved in Section 6.

**Remark 5.11. (on imperfect sampling)** In the description of our algorithms we will assume that we can perfectly sample from the distributions $\mu_\beta$. Of course, in applications we can only sample from distributions which are at a small variation distance $\delta$ from $\mu_\beta$.

Our algorithms will still work, as the following, standard, coupling trick shows. We can couple the biased distributions and the perfect distributions so that they differ with probability $\delta$. If we take $t$ samples total then, by union bound, with probability at least $1 - \delta t$ the algorithm with biased samplers will have the same output as the algorithm with perfect samplers.

**Remark 5.12. (on randomization)** The randomness in our algorithm will come from the procedures EST and IS-HEAVY. The failure probability parameter $\delta$ will be chosen very small so that during the execution of the algorithm no failures of EST and IS-HEAVY occur with high probability (formally, we use the union bound). Thus in the proof of correctness we will ignore the possibility of failure of these procedures and deal with the errors separately.

**Remark 5.13.** (**on binary search**) In our algorithm we will have to (approximately) find the right-most point in an interval $[a, b]$ which satisfies a given predicate $\Pi$. The predicate will be such that $\Pi(a)$ is true. We use the binary search on an interval $[a, b]$ in the following manner. If $\Pi(b)$ is true then we return $b$. Otherwise we set $\lambda = a$, $\rho = b$ and perform binary search until $\rho - \lambda \leq \varepsilon$, where $\varepsilon$ is the precision. Note that in the end we will have $\Pi(\rho)$ is false and $\Pi(\lambda)$ is true. We return $\lambda$.

We now give a detailed description of our algorithm for constructing the cooling schedule. Let $\delta'$ be the desired final error probability of our algorithm. We will call the procedures EST, IS-HEAVY, and FIND-HEAVY with the same value of $\delta$, which will be chosen as follows:

$$\delta = \frac{\delta'}{1600(\ln n)^2 (\ln A)^2}. \tag{62}$$

Let

$$s = \left\lceil (8/h) \ln \frac{1}{\delta} \right\rceil.$$

We will keep a set Bad of banned intervals which is initially empty.

Note it suffices to have the penultimate $\beta$ in the sequence be $\beta_{i-1} = \ln A$, since we can then set $\beta_i = \infty$ (see equation (10)). The algorithm for constructing the sequence works inductively. Thus, consider some starting $\beta_0$.

1. We first find an interval $I$ that is $h$-heavy at $\beta_0$ and is not banned. By generating $s$ samples from the distribution $\mu_{\beta_0}$ and taking the most frequently seen interval, we will successfully find an $h$-heavy interval with high probability (see Corollary 5.9 for the formal statement).

2. Let $w$ denote the width of $I$, i.e., $w = c - b$ where $I = [b, c]$. Our rough estimator (given by Lemma 5.10) only applies for $\beta_1 \leq \beta_0 + 1/w$ (by convention $1/0 = \infty$). Moreover, since we only need to reach a final inverse temperature of $\ln A$, let

$$L = \min\{\beta_0 + 1/w, \ln A\}.$$

Now we concentrate on constructing a cooling schedule within $(\beta_0, L]$.

3. Intuitively, we do binary search in the interval $[\beta_0, L]$ to find the maximum $\beta^*$ such that $\beta^*$ is $h$-heavy. We can use binary search because, by Lemma 5.5, the set of inverse temperatures for which an interval is heavy is an interval in $\mathbb{R}^+$. (More precisely, we do binary search in the interval $[\beta_0, L]$ with predicate IS-HEAVY$(I, \beta)$ and precision $\varepsilon = 1/(2n)$. We use the binary search procedure described in Remark 5.13.)

4. We now check if there is an "optimal" move within the interval

$$B' = (\beta_0, (\beta_0 + \beta^*)/2].$$

We want to find the maximum $\beta \in B'$ satisfying (52) for $u(\beta_0, \beta)$, or determine no such $\beta$ exists. Let $c_1 = e^2$ and $c_2 = 3 \cdot 10^6$ for (52). To find such a $\beta$, we do binary search and apply Lemma 5.10 to estimate the ratios $Z(2\beta - \beta_0)/Z(\beta)$ and $Z(\beta_0)/Z(\beta)$. Note for $\beta \in B'$ we have $2\beta - \beta_0 \in [\beta_0, \beta^*]$, hence, the interval $I$ is $h$-heavy at inverse temperatures $\beta_0, \beta$ and $2\beta - \beta_0$ and Lemma 5.10 applies.[1]

---

[1] More precisely, we perform binary search with predicate EST$(I, \beta_0, \beta) \cdot$EST$(I, 2\beta - \beta_0, \beta) \leq 2000$.

(a) If such a $\beta \in B'$ exists, then we set $\beta$ as the next inverse temperature and we repeat the algorithm starting from $\beta$. We refer to these steps as "optimal" moves.

(b) If no such $\beta$ exists, then we can reach the end of the interval $B$ as follows. There are two cases, either the interval was too wide for the application of Lemma 5.10, or the interval $I$ stops being heavy too soon. More precisely, either:

  i. If $\beta^* = L$, then we set $(\beta_0 + \beta^*)/2$ as the next inverse temperature. Moreover, if $\beta^* < \ln A$ we continue the algorithm starting from $\beta^*$; whereas if $\beta^* = \ln A$ we are done. We refer to these steps as "long" moves.

  ii. Otherwise, we add the following inverse temperatures to our schedule:

  $$\beta_0 + \frac{1}{2}\gamma, \beta_0 + \frac{3}{4}\gamma, \beta_0 + \frac{7}{8}\gamma, \ldots, \beta_0 + (1 - 2^{-t})\gamma, \beta_0 + \gamma,$$

  where $\gamma = \beta^* - \beta_0$ and $t = \lceil \ln \ln A \rceil$. We add the interval $I$ to the set of banned intervals Bad and continue the algorithm starting from $\beta^*$. We refer to these steps as "interval" moves since the interval $I$ will not be used by the algorithm again.

**Lemma 5.14** (Step 3). *Assume that no failures occurred. After step 3 of the algorithm, the interval $I$ is $h$-heavy for $\beta^*$. Moreover, if $\beta^* \neq L$, then the interval $I$ is not $8h$-heavy for $\beta^*$.*

**Lemma 5.15** (Step 4). *Assume that no failures occurred. Then after step 4 of the algorithm*

$$\frac{Z(\beta_0)Z(2\beta - \beta_0)}{Z(\beta)^2} \leq 3 \cdot 10^6. \tag{63}$$

*Moreover, if $\beta < (\beta^* + \beta_0)/2$, then*

$$\frac{Z(\beta_0)Z(2\beta - \beta_0)}{Z(\beta)^2} \geq e^2. \tag{64}$$

## 5.3 Bounding the length of the cooling schedule

We first estimate $|P|$, the number of intervals in $P$. It is used to bound the number of interval moves.

**Proof of Lemma 5.3:**
Let $i \in \{0, \ldots, n\}$. Suppose that the interval $I$ containing $i$ starts at $b$. Thus, by (55), the width of $I$ is $\lfloor b/\sqrt{\ln A} \rfloor$. Since $i$ is in $I$ we have

$$i \leq b + \lfloor b/\sqrt{\ln A} \rfloor \leq b(1 + 1/\sqrt{\ln A}) = b\frac{1 + \sqrt{\ln A}}{\sqrt{\ln A}}. \tag{65}$$

We can lower bound the width of the interval containing $i$ as follows (in the second inequality we use (65)):

$$\left\lfloor \frac{b}{\sqrt{\ln A}} \right\rfloor \geq \frac{b}{\sqrt{\ln A}} - 1 \geq \frac{i}{1 + \sqrt{\ln A}} - 1.$$

If for each $i \in \{0, \ldots, n\}$ we take the width $w$ of the interval containing $i$ and add up the $1/(w+1)$, we obtain the number of intervals. Thus, the total number of intervals is bounded as follows

$$|P| \leq 1 + \sum_{i=1}^{n} \frac{1 + \sqrt{\ln A}}{i} \leq 1 + (1 + \ln n)(1 + \sqrt{\ln A}) \leq 4\sqrt{\ln A} \ln n. \tag{66}$$

∎

We now bound the number of long moves.

**Lemma 5.16.** *Assume that no failures occurred during the algorithm. The number of "long" steps is bounded by $26\sqrt{\ln A} \ln n$.*

**Proof :**
At most one "long" move can have $L = \ln A$ (because the algorithm stops at the inverse temperature $\ln A$). Thus, we only need to estimate the number of "long" moves for which $L = \beta_0 + 1/w$.

Let $x_k$ be the total number of "long" moves for which the width of the interval $I$ was $k$. Let $k'$ be the largest $k$ such that $x_k$ is non-zero. Let $y_k = x_k$ for $k < k'$ and let $y_{k'} = x_{k'} - 1$.

Let $k \in \{1, \ldots, k'\}$. Let $t = (y_{k'} + y_{k'-1} + \cdots + y_k)$. After $t$ "long" moves the inverse temperature satisfies

$$\beta_0 \geq \sum_{i=k}^{k'} \frac{y_i}{2i} \tag{67}$$

($\beta_0$ would be equal to the right-hand side of (67) if we took the $t$ shortest "long" moves). Note that $x_{k'} + \cdots + x_k > t$, and, hence, we still have to make a long step with the width of the $h$-heavy interval $I$ at least $k$. This long step has to happen at an inverse temperature $\beta_0$, or higher.

We will need the following property - for any interval $[b, c] \in P$ of width $w = c - b$ and any $i \in [b, c]$ we have

$$i \geq b \geq w\sqrt{\ln A}. \tag{68}$$

This follows directly from (55) (since the chose the width $w$ to be $\lfloor b/\sqrt{\ln A} \rfloor$).

From (68) we have

$$\sum_{i \in I} a_i e^{-\beta_0 i} \leq A e^{-\beta_0 k \sqrt{\ln A}}. \tag{69}$$

Assume the left-hand side of (69) is $\leq h$. Then $I$ is not $h$-heavy for any $\rho \geq \beta_0$, since for $X \sim \mu_\rho$

$$\Pr(H(X) \in I) = \frac{\sum_{i \in I} a_i e^{-\rho i}}{Z(\rho)} \leq \frac{\sum_{i \in I} a_i e^{-\beta_0 i}}{Z(\rho)} \leq h,$$

in the last inequality we used $Z(\rho) \geq a_0 \geq 1$, which is true for any $\rho$. Thus, in the binary search in step 3) of the algorithm the IS-HEAVY will always report false and $\beta^*$ will be about $\beta_0 + 1/2n$ (more precisely $\beta^* \leq \beta_0 + 1/2n$). Hence $\beta^* < L$, which implies that a "long" move with $I$ of width $\geq k$ is impossible, a contradiction.

Thus, the left-hand side of (69) is $\geq h$, and hence

$$Ae^{-\beta k\sqrt{\ln A}} \geq h. \tag{70}$$

By combining (70) and (67) we obtain

$$\sum_{i=k}^{k'} \frac{y_i}{i} \leq 2\beta_0 \leq \frac{2}{k} \cdot \frac{\ln(A/h)}{\sqrt{\ln A}}. \tag{71}$$

Adding (71) for $k = 1, \ldots, k'$ we obtain

$$\sum_{i=1}^{k'} y_i \leq 2(1 + \ln n)\frac{\ln(A/h)}{\sqrt{\ln A}} \leq 4(\ln n)\left(\sqrt{\ln A} + \frac{\ln(1/h)}{\sqrt{\ln A}}\right). \tag{72}$$

By Lemma 5.3 and the definition of $h$ (equation (56)) we have

$$1/h \leq 32\sqrt{\ln A}\ln n.$$

and hence (using our assumptions (24)) we obtain

$$\ln(1/h) \leq 5\ln A.$$

The total number of long moves is thus bounded

$$2 + \sum_{i=1}^{k'} y_i \leq 2 + 24(\ln n)\sqrt{\ln A} \leq 26\sqrt{\ln A}\ln n.$$

∎

We now prove Theorem 5.1.
**Proof of Theorem 5.1:**
The number of "optimal" moves is bounded by

$$4\sqrt{(\ln A)\ln n}\ln\ln A, \tag{73}$$

see Corollary 4.4. Each "interval" move causes at most $s = 2\ln\ln A$ inverse temperatures to be output. Hence, by Lemma 5.3, the total number of inverse temperatures output by "interval moves" is bounded by

$$8\sqrt{\ln A}(\ln n)\ln\ln A. \tag{74}$$

Finally, the number of "long" moves is bounded by Lemma 5.16, and it is at most

$$26\sqrt{\ln A}\ln n. \tag{75}$$

The total number of moves is bounded by the sum of (73),(74),(75), which is bounded by $38\sqrt{\ln A}(\ln n)\ln\ln A$. This proves (50).

Let $T = 38\sqrt{\ln A}(\ln n)\ln\ln A$. The length of the output schedule is bounded by $T$ and hence every step of the algorithm is executed at most $T$ times. The binary search on step 3 starts with an interval of width at most $\ln A$ and works with precision $1/(4n)$. The total number of calls to EST is thus bounded by

$$2T \log_2(8n \ln A) \leq 8T(\ln n + \ln \ln A). \tag{76}$$

The total number of calls to IS-HEAVY is certainly bounded by (76), since the starting interval has width at most $\ln A$ and works with precision only $1/(2n)$. Finally, the number of calls to FIND-HEAVY is at most $T$.

Assuming perfect samples from the $\mu_\beta$, our algorithm can only fail inside EST, IS-HEAVY, and FIND-HEAVY. For each call this failure is bounded by $4\delta$ for the first two, and $|P|\delta$ for FIND-HEAVY (see Lemma 5.7, Lemma 5.10, and Corollary 5.9). By the union bound the total failure probability is bounded by

$$16T(\ln n + \ln \ln A)4\delta + T(4\sqrt{\ln A} \ln n)\delta \leq 60T\sqrt{\ln A}(\ln n)\delta$$
$$\leq 2400(\ln A)^2(\ln n)^2 \delta \leq \delta'/2.$$

where in the second to last inequality we used $\ln \ln A \leq \ln A$, and in the last inequality we used the definition of $\delta$ given by (62).

Of course requiring perfect samples is a too stringent requirement. Imperfect samples introduce one more source of error in our algorithm. As discussed in Remark 5.11 this is dealt with by a coupling argument. By our choice of the variation distance the imperfectness of samples manifests with probability at most $\delta'/2$.

The number of calls (per invocation) to the $\mu_\beta$ oracles made by any of the three procedures is

$$s = \left\lceil (8/h) \ln \frac{1}{\delta} \right\rceil \leq 512\sqrt{\ln A}(\ln n)\left(\ln 2400 + (2\ln \ln n) + (2\ln \ln A) + \ln \frac{1}{\delta'}\right)$$
$$\leq 10^4 \sqrt{\ln A}(\ln n)\left((\ln \ln n) + (\ln \ln A) + \ln \frac{1}{\delta'}\right). \tag{77}$$

Hence the total number of calls to the $\mu_\beta$ oracles is bounded by

$$Q \leq 20T((\ln n) + \ln \ln A)s \leq 10^7(\ln A)\big((\ln n) + \ln \ln A\big)^5 \ln \frac{1}{\delta'}.$$

∎

## 6  Leftover Proofs

**Proof of Lemma 4.2:**
We have

$$f(\beta) = \ln Z(\beta) = \ln \left(\sum_{i=0}^n a_i e^{-\beta i}\right).$$

Let $Y$ be the random variable defined by $Y = H(X)$ where $X \sim \mu_\beta$. We have

$$f'(\beta) = \frac{Z'(\beta)}{Z(\beta)} = \mathrm{E}(-Y) = -\mathrm{E}(Y).$$

Since the hamiltonian $H$ had values in the range $[0, n]$ we obtain parts (a) and (c) of the lemma.

Similarly

$$f''(\beta) = \frac{Z''(\beta)Z(\beta) - Z'(\beta)^2}{Z(\beta)^2} = \frac{Z''(\beta)}{Z(\beta)} - \left(\frac{Z'(\beta)}{Z(\beta)}\right)^2 = \mathrm{E}\left(Y^2\right) - \mathrm{E}\left(Y\right)^2 > 0,$$

by Jensen's inequality, proving part (b) of the lemma. ∎

**Proof of Lemma 5.7:**
Assume that $I$ is $4h$-heavy. Thus, the expected number of samples that fall inside $I$ is at least $4hs$. By the Chernoff bound (see, e.g., [5, Corollary 2.3]) it is very likely that the number of samples $X$ that fall inside $I$ greater than $2hs$. Formally,

$$\Pr\left(X \leq 2hs\right) \leq e^{-sh/8} \leq \delta. \tag{78}$$

Now assume that $I$ is not $h$-heavy. Thus, the expected number of samples that fall inside $I$ is at most $hs$. By the Chernoff bound,

$$\Pr\left(X \geq 2hs\right) \leq e^{-sh/8} \leq \delta. \tag{79}$$

∎

**Proof of Lemma 5.5:**
The interval $I$ is dense for $\beta = -\ln x$ if

$$0 \geq h\sum_{i=0}^{n} a_i x^i - \sum_{i \in I} a_i x^i =: g(x). \tag{80}$$

Note that $g(x)$ is a polynomial with at most 2 coefficient sign changes (i.e., looking at the coefficients sorted by the degree, the sign changes at most twice). Hence, by the Descartes' rule of signs it has at most 2 positive roots. Without loss of generality, we can assume that $n \notin I$ (otherwise we can "flip" the problem by $i \mapsto n - i$). Thus, $g(x)$ is positive at $x = \infty$ and hence the set of $x \in \mathbb{R}^+$ on which $g(x)$ is negative is an interval. Using the monotonicity of ln we obtain the result. ∎

**Proof of Lemma 5.10:**
Note that $\mathrm{E}\left(Y_k\right) = \Pr\left(X_k \in I\right) \geq h$. By the Chernoff bound for $k = 1, 2$ we have

$$\Pr\left(\frac{\mathrm{E}\left(Y_k\right)}{2} \leq U_k \leq 2\mathrm{E}\left(Y_k\right)\right) \geq 1 - 2e^{-hs/8},$$

and hence

$$\Pr\left(\frac{1}{4} \cdot \frac{\mathrm{E}\left(Y_1\right)}{\mathrm{E}\left(Y_2\right)} \leq \frac{U_1}{U_2} \leq 4 \cdot \frac{\mathrm{E}\left(Y_1\right)}{\mathrm{E}\left(Y_2\right)}\right) \geq 1 - 4e^{-hs/8}. \tag{81}$$

We have

$$\frac{\mathrm{E}\left(Y_1\right)}{\mathrm{E}\left(Y_2\right)} = \frac{Z(\beta_2)}{Z(\beta_1)} \cdot \frac{\sum_{i \in I} a_i e^{-\beta_1 i}}{\sum_{i \in I} a_i e^{-\beta_2 i}} = \frac{Z(\beta_2)}{Z(\beta_1)} \cdot e^{b(\beta_2 - \beta_1)} \cdot \frac{\sum_{i \in I} a_i e^{-\beta_2 i + (\beta_2 - \beta_1)(i - b)}}{\sum_{i \in I} a_i e^{-\beta_2 i}},$$

and therefore
$$\mathrm{e}^{-|\beta_1-\beta_2|(c-b)} \cdot \frac{Z(\beta_2)}{Z(\beta_1)} \leq \frac{\mathrm{E}(Y_1)}{\mathrm{E}(Y_2)} \cdot \mathrm{e}^{b(\beta_1-\beta_2)} \leq \mathrm{e}^{+|\beta_1-\beta_2|(c-b)} \cdot \frac{Z(\beta_2)}{Z(\beta_1)}. \tag{82}$$

Now combining (82), (81), and using assumption (59), the lemma follows. ∎

**Proof of Lemma 5.14:**
Is-Heavy($\beta^*, I$) reported $I$ as $h$-heavy for $\beta^*$. Assume that $\beta^* \neq L$. Then the binary search ended with interval $[\lambda, \rho]$ where $\lambda = \beta^*, \rho \leq \beta^* + 1/2n$. We have that Is-Heavy($\beta^*, I$) reported $I$ as not $4h$-heavy for $\beta^*$. Weight of an interval decreases by a factor of at most $\sqrt{\mathrm{e}}$ between $\lambda$ and $\rho$ and, hence, $I$ is not $8h$-heavy for $\alpha + \beta$. ∎

**Proof of Lemma 5.15:**
We use Est to refer to the procedure defined in Lemma 5.10. Since there were no failures, none of the calls to Est failed.

The predicate
$$\mathrm{Est}(I, \beta_0, x)\mathrm{Est}(I, 2x - \beta_0, x) \leq 2000 \tag{83}$$
was true for $x = \beta$. From (61), we obtain
$$\frac{Z(\beta_0)}{Z(\beta)} \frac{Z(2\beta - \beta_0)}{Z(\beta)} \leq (4\mathrm{e})^2 \mathrm{Est}(I, \beta_0, \beta)\mathrm{Est}(I, 2\beta - \beta_0, \beta),$$
and hence
$$\frac{Z(\beta_0)}{Z(\beta)} \frac{Z(2\beta - \beta_0)}{Z(\beta)} (4\mathrm{e})^2 2000 < 3 \cdot 10^6.$$
Assume that $\beta < (\beta^* + \beta_0)/2$. The binary search ended with an interval $[\lambda, \rho]$ where $\lambda = \beta$ and $\rho \leq \beta + 1/(4n)$. Using (61) we obtain
$$\frac{Z(\beta_0)}{Z(\rho)} \frac{Z(2\rho - \beta_0)}{Z(\rho)} \geq \frac{1}{(4\mathrm{e})^2} \mathrm{Est}(I, \beta_0, \rho)\mathrm{Est}(I, 2\rho - \beta_0, \rho).$$
The predicate (83) was false on $\rho$ and hence
$$\frac{Z(\beta_0)}{Z(\rho)} \frac{Z(2\rho - \beta_0)}{Z(\rho)} \geq \frac{2000}{(4\mathrm{e})^2}.$$
By Observation 3.1 we have $Z(2\beta - \beta_0) \geq Z(2\rho - \beta_0)$ and $Z(\beta) \leq Z(\rho)\mathrm{e}^{1/4}$.
$$\frac{Z(\beta_0)}{Z(\beta)} \frac{Z(2\beta - \beta_0)}{Z(\beta)} \geq \mathrm{e}^{-1/2} \frac{Z(\beta_0)}{Z(\rho)} \frac{Z(2\rho - \beta_0)}{Z(\rho)} \geq \mathrm{e}^2.$$
∎

# 7 Reversible Cooling Schedules for Warm Starts

In this section we show how to adapt the schedule generating algorithm to the setting of "warm starts", which often leads to faster sampling algorithms (see, e. g., [10, 11]).

This method reuses randomness to improve the overall running time. The downside is a slight dependence between random variables occurring in our algorithm. We will use the following notion of dependence.

**Definition 7.1.** Random variables $X, Y$ are $\kappa$-*independent* if for every (measurable) $A, B$ we have
$$|P(X \in A, Y \in B) - P(X \in A)P(Y \in B)| \leq \kappa.$$

We will need the following variant of Theorem 2.2, implicit in [11], which allows for slight dependence between its random variables.

**Theorem 7.2.** *Let $W = (W_1, \ldots, W_\ell)$ be a vector random variable. Let $\ell$ be a nonnegative integer, $K \geq 512\ell/\varepsilon^2$ and $\kappa \leq 2^{-20}\varepsilon^2/(K^5 \ell)$. Assume that*

- $W_i$ *is $\kappa$-independent from $(W_1, \ldots, W_{i-1})$, and*
- $\mathrm{E}\left(W_i^2\right)/\mathrm{E}\left(W_i\right)^2 \leq B$

*for $i \in [\ell]$. Let $\widehat{W} = W_1 \ldots W_\ell$. Let $S = (S_1, \ldots, S_\ell)$ be the average of $K$ samples from $W$. Let $\widehat{S} = S_1 S_2 \cdots S_\ell$. Then*
$$\Pr\left((1-\varepsilon)\mathrm{E}\left(\widehat{W}\right) \leq \widehat{S} \leq (1+\varepsilon)\mathrm{E}\left(\widehat{W}\right)\right) \geq 3/4.$$

We will use the following two notions of distance between probability distribution. For a pair of distributions $\nu$ and $\pi$ on a finite space $\Omega$, their *total variation distance* is defined as:
$$\|\nu - \pi\|_{\mathrm{TV}} = \frac{1}{2}\sum_{x \in \Omega} |\nu(x) - \pi(x)| = \max_{A \subset \Omega}(\nu(A) - \pi(A)).$$

In the applications section we will also need to consider $L^2$ *distance* defined by:
$$\left\|\frac{\nu}{\pi} - 1\right\|_{2,\pi}^2 = \mathrm{Var}_\pi(\nu/\pi) = \sum_{x \in \Omega} \pi(x)\left(\frac{\nu(x)}{\pi(x)} - 1\right)^2.$$

For a Markov chain $(X_t)$ with unique stationary distribution $\pi$ we will use the following result on the distance from stationarity after $t$ steps, starting from a "warm start" $\nu_0$. Let $\nu_t$ denote the distribution of $X_t, t \geq 0$. Let $\tau_2$ denote the inverse spectral gap, commonly known as the relaxation time, of the Markov chain. We will then use the following well-known fact, (see, e.g., [6], Theorem 5.6).

**Lemma 7.3.**
$$\|\nu_t - \pi\|_{\mathrm{TV}} \leq \exp(-t/2\tau_2)\left\|\frac{\nu_0}{\pi} - 1\right\|_{2,\pi}.$$

Let $\beta_0 = 0 < \cdots < \beta_\ell = \infty$ be a cooling schedule and let $\mu_i = \mu_{\beta_i}$ (for $i = 0, \ldots, \ell$). In our applications we will use the distribution from the previous round $(i)$ to serve as a warm start the current round $(i+1)$. For this we need that the "warm start" distribution $\mu_i$ is close to the distribution $\mu_{i+1}$, which is the stationary distribution for the current chain. We will use the $L_2$-notion of warm start, i.e., we will require inverse temperatures such that
$$\mathrm{Var}_{\mu_{i+1}}(\mu_i/\mu_{i+1}) = \left\|\frac{\mu_i}{\mu_{i+1}} - 1\right\|_{2,\mu_{i+1}} \tag{84}$$

is bounded. Then $\mu_i$ is a good "warm start" for $\mu_{i+1}$ and we can use Lemma 7.3 to upper bound the mixing time, obtaining a substantial improvement over the usual "cold start" bound (the saving comes from the fact that $\tau_2$ is often substantially smaller than the pessimistic "cold start" mixing time $\tau_{\mathrm{mix}}$).

A short calculation yields that the $L_2$ distance between distributions $\mu_i$ and $\mu_j$ can be expressed as a squared coefficient of variation of the variables arising in our algorithm. More precisely

$$\mathrm{Var}_{\mu_j}(\mu_i/\mu_j) = \frac{\mathrm{Var}\left(W_{\beta_j,\beta_i}\right)}{\mathrm{E}\left(W^2_{\beta_j,\beta_i}\right)} = \frac{Z(2\beta_i - \beta_j)Z(\beta_j)}{Z(\beta_i)^2} - 1 \leq \frac{Z(2\beta_i - \beta_j)Z(\beta_j)}{Z(\beta_i)^2}. \tag{85}$$

Note that for $j = i - 1$ the right-hand side of (85) becomes the right-hand side of the definition of $B$-Chebyshev cooling schedule (equation (6)). Thus for a $B$-Chebyshev cooling schedule

$$\mathrm{Var}_{\mu_i}(\mu_{i+1}/\mu_i) \leq B - 1. \tag{86}$$

The left-hand side of (86) is the left-hand side of (84) with the roles of $\mu_i$ and $\mu_{i+1}$ reversed. Thus, the condition that (84) be bounded is equivalent to saying that the schedule

$$\beta_\ell = \infty > \beta_{\ell-1} > \cdots > \beta_1 > 0 = \beta_0,$$

(i.e., the schedule in reverse) is a $B$-Chebyshev schedule for some constant $B$. This motivates the following definition.

**Definition 7.4.** Let $B > 0$ be a constant. Let $Z$ be a partition function. Let $\beta_0, \ldots, \beta_\ell$ be a sequence of inverse temperatures such that $0 = \beta_0 < \beta_1 < \cdots < \beta_\ell = \infty$. The sequence is called a *reversible $B$-Chebyshev cooling schedule* for $Z$ if

$$\frac{Z(2\beta_{i+1} - \beta_i)Z(\beta_i)}{Z(\beta_{i+1})^2} \leq B, \tag{87}$$

and

$$\frac{Z(2\beta_i - \beta_{i+1})Z(\beta_{i+1})}{Z(\beta_i)^2} \leq B, \tag{88}$$

for all $i = 0, \ldots, \ell - 1$.

Given a $B$-Chebyshev cooling schedule of length $\ell$ it is relatively easy to produce a reversible $B$-Chebyshev cooling schedule. We do so at the expense of an extra $O((\ln n) + \ln \ln A)$ factor in the length of the schedule. We will augment each interval $[\beta_i, \beta_{i+1}]$, $i = 0, \ldots, \ell - 2$ by careful initial steps. Let $t$ be the largest integer such that $2^t/n \leq \beta_{i+1} - \beta_i$. Note that $t = O((\ln n) + \ln \ln A)$. We insert the following inverse temperatures between $\beta_i$ and $\beta_{i+1}$

$$\beta_i + 1/n, \beta_i + 2/n, \beta_i + 4/n, \ldots, \beta_i + 2^t/n. \tag{89}$$

For $\beta = \beta_i$ and $\beta' = \beta_i + 1/n$ we have, by Lemma 4.2:

$$\frac{Z(2\beta - \beta')Z(\beta')}{Z(\beta)^2} \leq \mathrm{e}.$$

For $\beta = \beta_i + 2^j/n$ and $\beta' = \beta_i + 2^{j+1}/n$ we have $2\beta - \beta' = \beta_i$ and $\beta' \leq \beta_{i+1}$. Hence

$$\frac{Z(2\beta - \beta')Z(\beta')}{Z(\beta)^2} = \frac{Z(\beta_i)Z(2\beta - \beta_i)}{Z(\beta)^2} \leq \frac{Z(\beta_i)Z(2\beta_{i+1} - \beta_i)}{Z(\beta_{i+1})^2} \leq B,$$

since we started with a $B$-Chebyshev cooling schedule. For $\beta = \beta_i + 2^t/n$ and $\beta' = \beta_{i+1}$ the argument is the same.

**Theorem 7.5.** *Let $Z$ be a partition function. Suppose that for every inverse temperature $\beta$ we have a Markov chain $M_\beta$ with stationary distribution $\mu_\beta$. Assume that the relaxation time of all the $M_\beta$ chains is uniformly bounded by $\tau_2$. Assume that we can directly sample from $\mu_0$.*

*With probability at least $1-\delta'$, we can produce a reversible $B$-Chebyshev cooling schedule $\beta_0 = 0 < \beta_1 < \cdots < \beta_{\ell-1} < \beta_\ell = \infty$, for $B = 3 \cdot 10^6$, with*

$$\ell \leq 38\sqrt{\ln A}(\ln n)(\ln \ln A)((\ln n) + \ln \ln A).$$

*The algorithm uses at most*

$$Q \leq 10^7 (\ln A)\big((\ln n) + \ln \ln A\big)^5 \tau_2 \ln \frac{1}{\delta'}$$

*steps of the $M_\beta$ chains.*

**Proof :**
We will run the original algorithm and the augment the schedule using (89). To facilitate warm starts we will use the non-adaptive cooling schedule

$$\beta'_0 = 0 < \beta'_1 < \cdots < \beta'_{\ell'} = \infty \tag{90}$$

of [1] (equation (11) in this paper). We start with random sample at the inverse temperature 0, run $M_{\beta'_1}$ for $\tau_2$ steps, then $M_{\beta'_2}$ for $\tau_2$ steps and so on. This way we obtain warm starts for all inverse temperatures in the schedule (90). Note that we made $O(\tau_2(\ln n) \ln \ln A)$ steps of the chains so far.

Now we can run algorithm PRINT-COOLING-SCHEDULE and spend only $\tau_2$ steps to generate a random sample at any inverse temperature $\beta$. We use the closest inverse temperature in (90) as a warm start. ∎

Combining Theorem 7.2 with Theorem 7.5 we obtain.

**Corollary 7.6.** *Let $Z$ be a partition function. Let $\varepsilon > 0$ be the desired precision. Suppose that for every inverse temperature $\beta$ we have a Markov chain $M_\beta$ with stationary distribution $\mu_\beta$. Assume that the relaxation time of all the $M_\beta$ chains is uniformly bounded by $\tau_2$. Assume that we can directly sample from $\mu_0$. Using*

$$\tau_2 \frac{10^{10}}{\varepsilon^2}(\ln A)\big((\ln n) + \ln \ln A\big)^7 \ln \frac{10^8 (\ln A)\big((\ln n) + \ln \ln A\big)^7}{\varepsilon^2}$$

*steps of the $M_\beta$ chains we can obtain a random variable $\widehat{S}$ such that*

$$P\big((1-\varepsilon)Z(\infty) \leq \widehat{S} \leq (1+\varepsilon)Z(\infty)\big) \geq 3/4.$$

# 8  Applications

We detail several specific applications of our work: matchings, Ising model, colorings and independent sets. To simplify the comparison of our results with previous work and since we have not optimized polylogarithmic factors in our work, we use $O^*()$ notation which hides polylogarithmic terms and the dependence on $\epsilon$. Our cooling schedule results in a savings of a factor of $O^*(n)$ in the running time for all of the approximate counting problems considered here.

## 8.1  Matchings

We first consider the problem of generating a random matching of an input graph $G = (V, E)$. Let $\lambda = \exp(-1/\beta)$ and let $\Omega$ denote the set of matchings of $G$. For $M \in \Omega$, let $w(M) = \lambda^{|M|}$ (where $0^0 = 1$). The Gibbs distribution is then $\mu(M) = w(M)/Z$ where $Z = \sum_{M'} w(M')$. Note, for $\beta = \infty$ (i.e., $\lambda = 1$), $\mu$ is uniform over $\Omega$, whereas for $\beta = 0$ (i.e., $\lambda = 0$), $Z = 1$ since the empty set is the only matching with positive weight,

Consider the following ergodic Markov chain with stationary distribution $\mu$. Let $X_0 \in \Omega$ where $w(X_0) > 0$. From $X_t \in \Omega$,

- Choose $e = (u, v)$ uniformly at random from $E$.

- Set
$$X' = \begin{cases} X_t \setminus e & \text{if } e \in X_t \\ X_t \cup e & \text{if } u \text{ and } v \text{ are unmatched in } X_t \\ X_t \cup e \setminus (v, w) & \text{if } u \text{ is unmatched in } X_t \text{ and } (v, w) \in X_t \\ X_t \cup e \setminus (u, z) & \text{if } v \text{ is unmatched in } X_t \text{ and } (u, z) \in X_t \\ X_t & \text{otherwise} \end{cases}$$

- Let $X_{t+1} = X'$ with probability $\min\{1, w(X')/w(X_t)\}/2$, and otherwise set $X_{t+1} = X_t$.

Jerrum and Sinclair [8] proved that the above Markov chain has relaxation time $\tau_2 = O(nm)$ (see [6] for the claimed upper bound).

Since $A \leq n!2^n$, using Theorem 7.5 we obtain a cooling schedule of length $\ell = O(\sqrt{n}\log^4 n)$. In contrast, the previous best schedule was presented by [1] which had length $O(n \log^2 n)$. Thus, we save a factor of $O^*(n)$ in the running time for approximating $Z$. Applying Corollary 7.6 we obtain the following result.

**Corollary 8.1.** *For any $G = (V, E)$, for all $\varepsilon > 0$, let $\mathcal{M}(G)$ denote the set of matchings of $G$. We can compute an estimate $EST$ such that:*

$$EST(1 - \varepsilon) \leq |\mathcal{M}(G)| \leq EST(1 + \varepsilon)$$

*with probability $\geq 3/4$ in time $O(n^2 m \varepsilon^{-2} \log^7 n) = O^*(n^2 m)$.*

Recall, the error probability $3/4$ can be replaced by $1 - \delta$, for any $\delta > 0$, at the expense of an extra factor of $O(\log(1/\delta))$ in the running time.

## 8.2 Spin Systems

Spin systems are a general class of statistical physics models where our results apply. We refer the reader to [13, 16] for an introduction to spin systems. The examples we highlight here are well-studied examples of spin systems. Recall, the mixing time of a Markov chain is the number of transitions (from the worst initial state) to reach within variation distance $\leq \delta$ of the stationary distribution, where $0 < \delta < 1$. The following results follow in a standard way from the stated mixing time result combined with Corollary 5.2.

**Colorings:** For a graph $G = (V, E)$ with maximum degree $\Delta$ we are interested in approximating the number of $k$-colorings of $G$. Here, we are coloring the vertices using a palette of $k$ colors so that adjacent vertices receive different colors. This problem is also known as the zero-temperature (thus $\beta = \infty$) anti-ferromagnetic Potts model. The simple single-site update Markov chain known as the Glauber dynamics is ergodic with unique stationary distribution uniform over all $k$-colorings whenever $k \geq \Delta + 2$ There are various regions where fast convergence of the Glauber dynamics is known, we refer the interested reader to [4] for a survey. For concreteness we consider the result of Jerrum [7] who proved that the Glauber dynamics has mixing time $O(kn \log(n/\delta))$ whenever $k > 2\Delta$. Moreover, his proof easily extends to any non-zero temperature. (Recall, the mixing time of a Markov chain is the number of steps so that, from the worst initial state, we are within variation distance $\leq \delta$ of the stationary distribution.) Since $A = k^n$, using Corollary 5.2 we obtain the following result.

**Corollary 8.2.** *For all $k > 0$, any graph $G = (V, E)$ with maximum degree $\Delta$, let $\Omega(G)$ denote the set of $k$-colorings of $G$. For all $\varepsilon > 0$, whenever $k > 2\Delta$, we can compute an estimate EST such that:*

$$EST(1 - \varepsilon) \leq |\Omega(G)| \leq EST(1 + \varepsilon)$$

*with probability $\geq 3/4$ in time $O(kn^2 \varepsilon^{-2} \log^6 n) = O^*(n^2)$.*

In comparison, the previous bound [1] required $O^*(n^3)$ time (and Jerrum [7] required $O^*(nm^2)$ time).

**Ising model:** There are extensive results on sampling from the Gibbs distribution and approximating the partition function of the (ferromagnetic) Ising model. We refer the reader to [13] for background and a survey of results. We consider a particularly well-known result. For the Ising model on an $\sqrt{n} \times \sqrt{n}$ 2-dimensional grid, Martinelli and Olivieri [14] proved that the Glauber dynamics (i.e., single-site update Markov chain) has mixing time $O(n \log(n/\delta))$ for all $\beta > \beta_c$ where $\beta_c$ is the critical point for the phase transition between uniqueness and non-uniqueness of the infinite-volume Gibbs measure. In this setting, we have $A = 2^n$ and, hence, we obtain the following result.

**Corollary 8.3.** *For the Ising model on a $\sqrt{n} \times \sqrt{n}$ 2-dimensional grid, let $Z(\beta)$ denote the partition function at inverse temperature $\beta > 0$. For all $\varepsilon > 0$, for all $\beta > \beta_c$, we can compute an estimate EST such that:*

$$EST(1 - \varepsilon) \leq Z(\beta) \leq EST(1 + \varepsilon)$$

*with probability $\geq 3/4$ in time $O(n^2 \varepsilon^{-2} \log^6 n) = O^*(n^2)$.*

**Independent Sets:** Given a fugacity $\lambda > 0$ and a graph $G = (V, E)$ with maximum degree $\Delta$, we are interested in computing

$$Z_G(\lambda) = \sum_{\sigma \in \Omega} \lambda^{|\sigma|},$$

where $\Omega$ is the set of independent sets of $G$. This is known as hard-core lattice gas model. In [15], it was proved that the Glauber dynamics for sampling from the distribution corresponding to $Z_G(\lambda)$ has $O(n \log(n/\delta))$ mixing time whenever $\lambda < 2/(\Delta - 2)$. As a consequence, we obtain the following result.

**Corollary 8.4.** *For any graph $G = (V, E)$ with maximum degree $\Delta$, For all $\varepsilon > 0$, for any $\lambda < 2/(\Delta - 2)$, we can compute an estimate EST such that:*

$$EST(1 - \varepsilon) \leq Z_G(\lambda) \leq EST(1 + \varepsilon)$$

*with probability $\geq 3/4$ in time $O(n^2 \varepsilon^{-2} \log^6 n) = O^*(n^2)$.*

Note, Weitz [17] has an alternative approach for this problem. His approach approximates $Z_G(\lambda)$ directly (without using sampling) and holds for a larger range of $\lambda$ (though $\Delta$ is required to be constant).

## 9 Discussion

An immediate question is whether these results extend to estimating the permanent of a 0/1 matrix. Our current adaptive scheme works assuming a sampling subroutine that can produce samples at any given temperature (from a warm start). The permanent algorithm of [9] also requires a set of $n^2 + 1$ weights to produce samples from a given temperature. These weights are computed from $n^2 + 1$ partition functions and it appears that a schedule of length $\Omega(n)$ is necessary if one considers all $n^2 + 1$ partition functions simultaneously. In fact, this is the case for the standard bad example of a chain of boxes (or a chain of hexagons as illustrated in Figure 2 of [9]).

## References

[1] I. Bezáková, D. Štefankovič, V. Vazirani, and E. Vigoda, Accelerating simulated annealing for combinatorial counting. In *Proceedings of the 17th Annual ACM-SIAM Symposium on Discrete Algorithms* (SODA), 900–907, 2006.

[2] M. E. Dyer and A. Frieze. Computing the volume of a convex body: a case where randomness provably helps. In *Proceedings of AMS Symposium on Probabilistic Combinatorics and Its Applications*, 123–170, 1991.

[3] M.E. Dyer, A.M. Frieze and R. Kannan, A random polynomial time algorithm for approximating the volume of convex bodies. *Journal of the ACM*, 38(1):1–17, 1991.

[4] A. Frieze and E. Vigoda, A survey on the use of Markov chains to randomly sample colorings. *Combinatorics, Complexity and Chance*, Oxford University Press, to appear, 2007.

[5] S. Janson, T. Łuczak, and A. Ruciński, *Random Graphs*. Wiley-Interscience Series in Discrete Mathematics and Optimization, 2000.

[6] M. Jerrum, *Counting, sampling and integrating: algorithms and complexity*. Lectures in Mathematics, Birkhäuser Verlag, 2003.

[7] M. Jerrum, A very simple algorithm for estimating the number of $k$-colorings of a low-degree graph. *Random Structures and Algorithms*, 7(2):157–165, 1995.

[8] M. Jerrum and A. Sinclair, Approximating the permanent. *SIAM Journal on Computing*, 18:1149–1178, 1989.

[9] M. Jerrum, A. Sinclair, and E. Vigoda, A polynomial-time approximation algorithm for the permanent of a matrix with non-negative entries. *Journal of the ACM*, 51(4):671-697, 2004.

[10] R. Kannan, L. Lovász, and M. Simonovits, Random walks and an $O^*(n^5)$ volume algorithm for convex bodies. *Random Structures and Algorithms* 11, 1–50, 1997.

[11] L. Lovász and S. Vempala, Simulated annealing in convex bodies and an $O^*(n^4)$ volume algorithm. *Journal of Computer and System Sciences*, 72(2):292-417, 2006.

[12] L. Lovász and S. Vempala, Fast algorithms for logconcave functions: sampling, rounding, integration and optimization. In *Proceedings of the 47th Annual IEEE Symposium on Foundations of Computer Science* (FOCS), 2006.

[13] F. Martinelli, Relaxation times of Markov chains in statistical mechanics and combinatorial structures, *Encyclopedia of Mathematical Sciences*, Vol. 110, Springer, 2003.

[14] F. Martinelli and E. Olivieri, Approach to equilibrium of Glauber dynamics in the one phase region I: The attractive case, *Communications in Mathematical Physics*, 161:447-486, 1994.

[15] E. Vigoda, A note on the Glauber dynamics for sampling independent sets. *Electronic Journal of Combinatorics*, 8(1), 2001.

[16] D. Weitz, Mixing in time and space for discrete spin systems, Ph.D. thesis, U.C. Berkeley, May 2004. Available from http://dimacs.rutgers.edu/∼dror/thesis/thesis.pdf

[17] D. Weitz, Counting independent sets up to the tree threshold. In *Proceedings of the 38th Annual ACM Symposium on Theory of Computing* (STOC), 140–149, 2006.

# 10 Appendix

## 10.1 Algorithm Pseudocode

---

    **input** : A black-box sampler for $X \sim \mu_\beta$ for any $\beta \geq 0$,
          starting inverse temperature $\beta_0$.
    **output**: A cooling schedule for $Z$.

    Bad $\leftarrow \emptyset$
    print $\beta_0$
    **if** $\beta_0 < \ln A$ **then**
1        $I \leftarrow $ Find-Heavy$(\beta_0, \text{Bad})$
2        $w \leftarrow$ the width of $I$
        $L \leftarrow \min\{\beta_0 + 1/w, \ln A\}$;            (where $1/0 = \infty$)
3        $\beta^* \leftarrow$ binary search on $\beta^* \in [\beta_0, L]$
              with precision $1/(2n)$,
              using predicate Is-Heavy$(\beta^*, I)$
4        $\beta \leftarrow$ binary search on $\beta \in [\beta_0, (\beta^* + \beta_0)/2]$
              with precision $1/(4n)$,
              using predicate Est$(I, \beta_0, \beta)\cdot$Est$(I, 2\beta - \beta_0, \beta) \leq 2000$
       **if** $\beta < (\beta^* + \beta_0)/2$ **then**
           Print-Cooling-Schedule$(\beta)$               ("optimal" move)
       **else**
           **if** $\beta = L$ **then**
               Print-Cooling-Schedule$(\beta)$               ("long" move)
           **else**
               $\gamma \leftarrow (\beta^* - \beta_0)/2$
               print $\beta_0 + \gamma, \beta_0 + (3/2)\gamma, \beta_0 + (7/4)\gamma, \ldots, \beta_0 + (2 - 2^{-\lceil \ln \ln A \rceil})\gamma$
               Bad $\leftarrow$ Bad $\cup\, I$
               Print-Cooling-Schedule$(\beta^*)$               ("interval" move)
           **end**
       **end**
    **else**
       print $\infty$
    **end**

**Algorithm 1**: Print-Cooling-Schedule